\documentclass[letterpaper]{article} 
\usepackage{aaai2026}  
\usepackage{times}  
\usepackage{helvet}  
\usepackage{courier}  

\usepackage[hyphens]{url}  
\usepackage{graphicx} 
\usepackage{natbib}  
\usepackage{caption} 
\usepackage{algorithmic}
\usepackage{amssymb} %
\usepackage{amsmath}
\usepackage{multirow}
\usepackage{booktabs}  
\usepackage{xspace}    
\usepackage{graphicx}  
\usepackage{subfigure}
\usepackage{amsthm}

\usepackage{url}
\usepackage{enumitem}

\usepackage{xcolor}
\usepackage{pifont}
\usepackage{makecell}
\usepackage{newfloat}
\usepackage{listings}
\usepackage[linesnumbered,ruled,vlined]{algorithm2e}

\DeclareCaptionStyle{ruled}{labelfont=normalfont,labelsep=colon,strut=off} 
\title{Task-Aware Retrieval Augmentation for Dynamic Recommendation}
\author{
  Zhen Tao$^{\bigstar \blacklozenge }$\equalcontrib, 
  Xinke Jiang$^{\diamondsuit}$\equalcontrib, 
  Qingshuai Feng$^{\bigstar}$, 
  Haoyu Zhang$^{\clubsuit}$, 
  Lun Du$^{\spadesuit}$, \\
  Yuchen Fang$^{\heartsuit}$, 
  Hao Miao$^{\triangle}$, 
  Bangquan Xie$^{\bigstar}$, 
  \textbf{Qingqiang Sun$^{\bigstar}$}\thanks{Corresponding author.} \\
  \normalfont{$^{\bigstar}$Great Bay University} 
  \normalfont{$^{\blacklozenge}$Nanjing University} 
  \normalfont{$^{\diamondsuit}$No Affiliation} \\
  \normalfont{$^{\clubsuit}$City University of Hong Kong} 
  \normalfont{$^{\spadesuit}$Ant Research} \\
  \normalfont{$^{\heartsuit}$University of Electronic Science and Technology of China} \\
  \normalfont{$^{\triangle}$Hong Kong Polytechnic University} \\
  \includegraphics[height=0.83em]{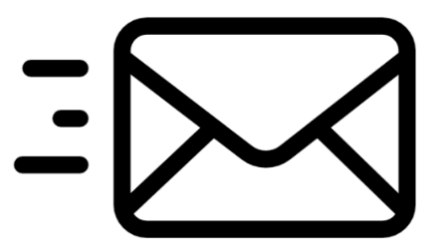} 
  {zhentao.tz@gmail.com, qqsun@gbu.edu.cn} \\
}

\usepackage{bibentry}

\begin{document}

\maketitle

\begin{abstract}
Dynamic recommendation systems aim to provide personalized suggestions by modeling temporal user-item interactions across time-series behavioral data. Recent studies have leveraged pre-trained dynamic graph neural networks (GNNs) to learn user-item representations over temporal snapshot graphs. However, fine-tuning GNNs on these graphs often results in generalization issues due to temporal discrepancies between pre-training and fine-tuning stages, limiting the model's ability to capture evolving user preferences. To address this, we propose TarDGR, a task-aware retrieval-augmented framework designed to enhance generalization capability by incorporating task-aware model and retrieval-augmentation. Specifically, TarDGR introduces a Task-Aware Evaluation Mechanism to identify semantically relevant historical subgraphs, enabling the construction of task-specific datasets without manual labeling. It also presents a Graph Transformer-based Task-Aware Model that integrates semantic and structural encodings to assess subgraph relevance. During inference, TarDGR retrieves and fuses task-aware subgraphs with the query subgraph, enriching its representation and mitigating temporal generalization issues. Experiments on multiple large-scale dynamic graph datasets demonstrate that TarDGR consistently outperforms state-of-the-art methods, with extensive empirical evidence underscoring its superior accuracy and generalization capabilities.

\end{abstract}

\section{Introduction}
Dynamic recommendation systems aim to generate personalized suggestions by modeling how user-item interactions evolve over time through rich temporal behavior data~\cite{zhu2021popularity, lv2024semantic, yang2023generic,yu2024dygprompt,  chen2025fairdgcl}. To capture such temporal dynamics, recent approaches have adopted Graph Neural Networks (GNNs)~\cite{yu2024non, yu2023hgprompt,jiang2025time,jiang2023incomplete} that learn user-item representations over sequences of time-evolving snapshot graphs~\cite{yang2024graphpro, jiang2024ragraph, yu2024dygprompt}. These GNN-based methods typically follow a pretraining–finetuning paradigm, where models are first trained on historical graphs to learn transferable structural patterns, and subsequently fine-tuned on recent temporal graphs to adapt to evolving user behavior. However, despite their success, these models often suffer from \textit{generalization issues} when fine-tuned on new temporal graphs, due to temporal discrepancies between pre-training and fine-tuning interaction graphs~\cite{cong2024generalization, lu2024temporal, tao2025dynamic, yu2023generalized,yu2025gcot, liang2025topology}. As the temporal context shifts and user interests evolve, previously learned patterns may no longer align with the current data distribution, limiting the model’s ability to deliver accurate recommendations for future interactions. 

\begin{figure}[t]
    \centering
    \includegraphics[width=0.99\linewidth]{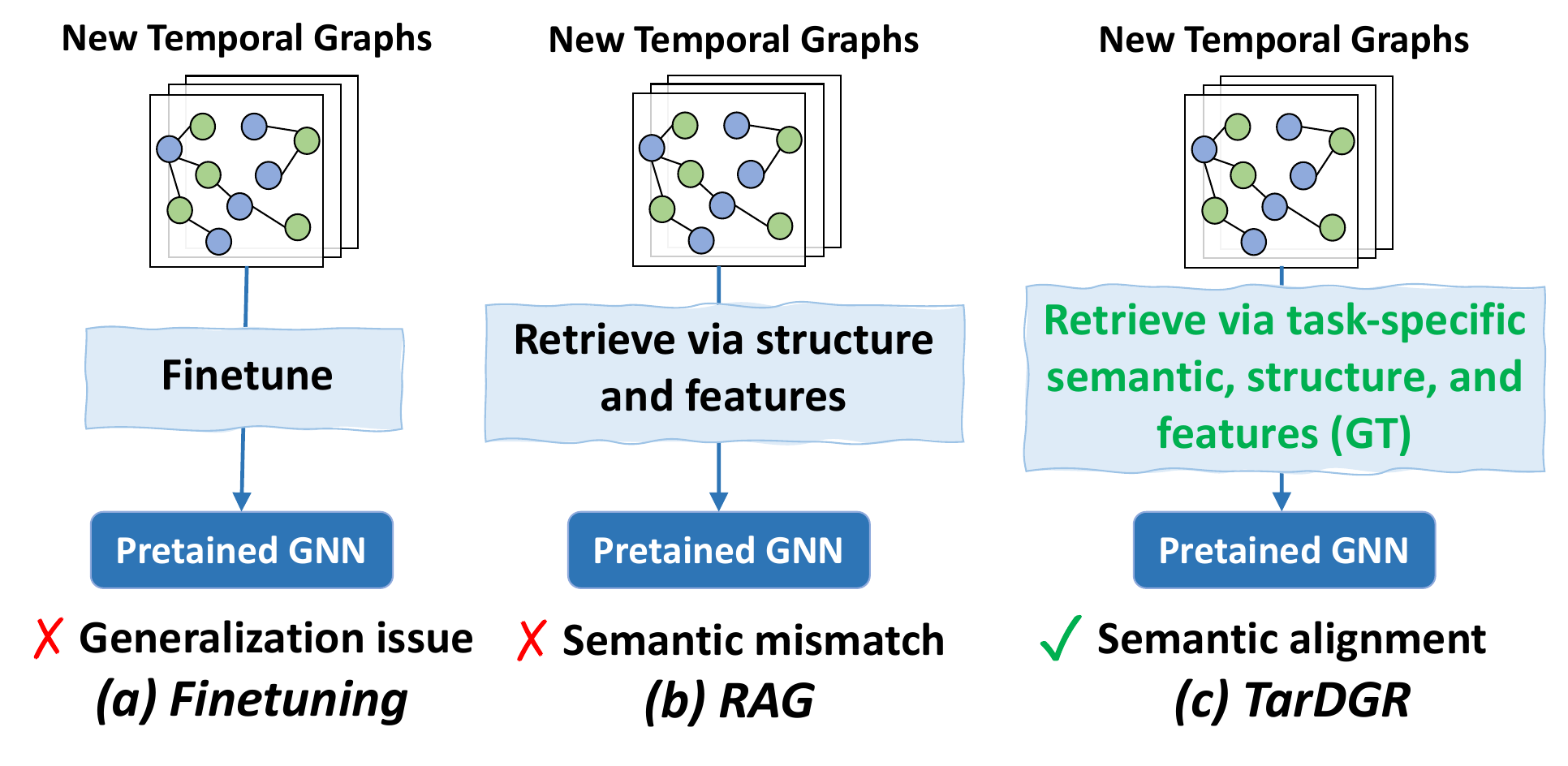}
    \caption{Comparison of current methods with TarDGR in dynamic graph recommendation.}
    \label{fig:comparison}
    \vspace{-0.7cm}
\end{figure}

\begin{figure}[t]
    \centering
    \includegraphics[width=0.99\linewidth]{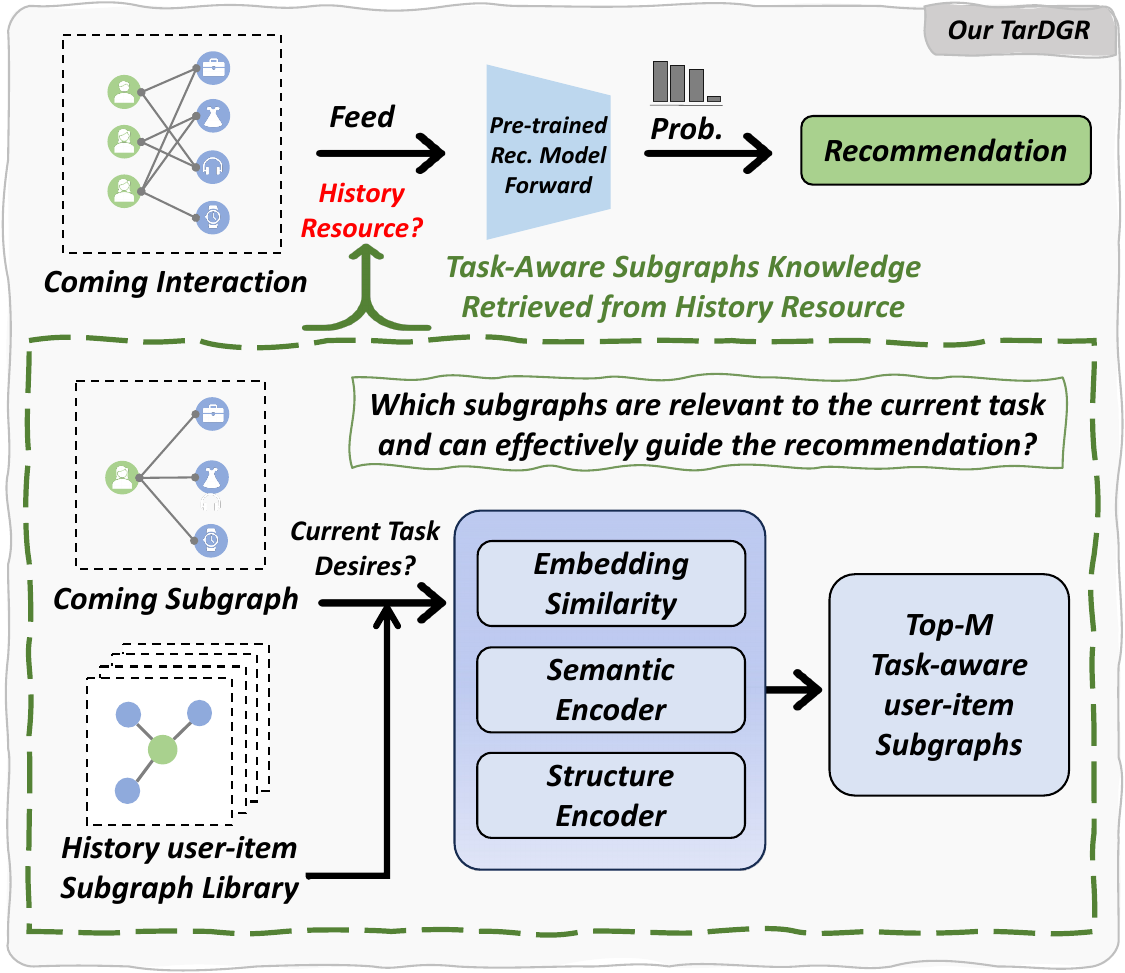}
    \caption{Task-Aware Retrieval Recommendation.}
    \label{fig:intro}
    \vspace{-0.5cm}
\end{figure}

Recent advances in Retrieval-augmented Generation (RAG) techniques have shown promise in addressing generalization issues by incorporating dynamic retrieval mechanisms, significantly enhancing model capabilities without requiring parameter updates~\cite{wu2024retrieval, zheng2025retrieval, parashar2024neglected,jiang2025hykge}. For dynamic recommendation tasks, providing appropriate context for predicting time-sensitive recommendation representations is crucial for improving generalization capability. While existing works have introduced external graph data through structural and feature similarity-based retrieval, they often overlook the semantic task relevance between retrieval and query graphs~\cite{jiang2024ragraph}. This limitation is particularly problematic for recommendation tasks, where structurally similar subgraphs may not be beneficial when semantic inconsistency exists among nodes. 
A comparison of these paradigms is illustrated in Figure~\ref{fig:comparison}, highlighting their respective temporal adaptation strategies and limitations. To fully realize the potential of retrieval-augmented recommendation in dynamic environments, we must overcome the following key challenges:
\textbf{C1. How to effectively identify task-relevant subgraphs for recommendation?} Current retrieval-based approaches for graph models rely primarily on structural and feature similarity when retrieving historical subgraphs. They overlook semantic task relevance between retrieval and query graphs, assuming all structurally similar subgraphs provide equal value. For recommendation tasks, this assumption proves problematic as structurally similar subgraphs may contribute little or even negatively when semantic inconsistency exists. Existing frameworks lack mechanisms to evaluate whether retrieved subgraphs actually benefit the specific recommendation task at hand.

\textbf{C2. How to enable models to understand task-specific requirements without manual annotation?} In the realm of graph recommendation, models struggle to interpret what kind of subgraph information best serves the current query recommendation graph. Unlike language model research where task-awareness can be incorporated through dataset construction~\cite{qiao2025agentic,sun2025vision,liu2024datasets}, graph data's inherent complexity makes manual dataset creation nearly infeasible. Without properly understanding task requirements, models cannot effectively transfer knowledge from historical data to new temporal contexts, resulting in suboptimal recommendations in dynamic environments.

To address these challenges, we propose \textbf{TarDGR} (\textbf{\underline{T}}ask-\textbf{\underline{A}}ware \textbf{\underline{R}}etrieval Augmentation for \textbf{\underline{D}}ynamic \textbf{\underline{G}}raph \textbf{\underline{R}}ecommendation), 
a novel framework that enhances generalization capability by incorporating task-aware model and retrieval-augmentation. As illustrated in Figure~\ref{fig:intro}, TarDGR identifies and injects task-specific knowledge into the query which enriches the representation.
First, to tackle \textbf{C1}, TarDGR introduces a \textit{Task-Aware Evaluation Mechanism} that automatically identifies semantically relevant historical resource subgraphs by evaluating how their integration affects similarity with positive recommendation samples. This enables the construction of task-specific datasets without manual annotations, defining clear criteria for ``task-beneficial" and ``task-harmful" subgraphs. Second, addressing \textbf{C2}, TarDGR presents a \textit{Graph Transformer-based Task-Aware Model} that combines semantic and structural encodings to evaluate the relevance of subgraphs in relation to current recommendation task requirements. 
During inference, the model retrieves and fuses task-relevant subgraphs with the query subgraph, thereby enhancing its representation ability and mitigating generalization issues induced by temporal shifts. 
We summarize our contributions as follows:
\begin{itemize}
\item We propose a task-aware retrieval-augmented framework for dynamic graph recommendation, offering the first exploration of task-awareness in graph learning for recommendation.
\item We introduce a novel automatic task-aware evaluation framework that distinguishes task-beneficial and task-harmful subgraphs, enabling the construction of task-aware datasets without manual annotations.
\item We design a Graph Transformer-based Task-Aware Model that effectively captures subgraph relevance by integrating semantic and structural encodings, enabling more accurate relevance estimation and robust knowledge transfer under temporal shifts. 
\item We apply the task-aware retrieval-augmentation technique to dynamic graph recommendation systems and demonstrate that TarDGR consistently outperforms state-of-the-art methods across three real-world datasets.
\end{itemize}

\section{Related Works}
\subsection{Dynamic Recommendation}

Dynamic recommendation has been addressed through sequential models such as BERT4Rec~\cite{sun2019bert4rec} and DCRec~\cite{yang2023debiased}, which rely on fixed historical sequences without explicitly modeling temporal graph structures. To better capture structural dynamics, dynamic graph neural networks (DGNNs) have emerged, including EvolveGCN~\cite{pareja2020evolvegcn}, ROLAND~\cite{you2022roland}, and WinGNN~\cite{zhu2023wingnn}, which explicitly model structural evolution over time. In parallel, the pretraining–finetuning paradigm has proven effective for transferring structural knowledge in graph learning~\cite{yu2025uniform, yu2025non, yu2024survey}. GraphPro~\cite{yang2024graphpro} extends this to dynamic recommendation by incorporating temporal prompts during pretraining and fine-tuning, achieving promising performance. However, significant temporal shifts between pretraining and fine-tuning snapshots can lead to generalization issues, ultimately limiting the model’s ability to adapt to evolving user preferences.

\subsection{Retrieval-Augmented Generation}

RAG enhances pre-trained language models by retrieving external knowledge to construct informative context for downstream tasks~\cite{rag, lewis2021retrievalaugmented}. These systems typically retrieve documents or entities from large corpora to guide generation, improving factuality and interpretability~\cite{sarthi2024raptor, gao2022precise}. RAG has been successfully extended to various modalities, including vision, code, audio, and video~\cite{zheng2025retrieval, yang2025empirical, xue2024retrieval, singh2025agentic}. In the graph domain, RAG has been applied to knowledge graphs by leveraging node-level textual features~\cite{xu2024generate}. Recent works like RAGRAPH~\cite{jiang2024ragraph} propose plug-and-play retrieval to augment pre-trained GNNs. However, these methods do not consider task-specific relevance during retrieval. 
As a result, semantically misaligned subgraphs may be introduced, degrading the quality of the downstream recommendation. 
Our approach addresses this gap by incorporating task-aware retrieval mechanisms.

\section{Preliminaries}
\paragraph{Problem Formulation}
We model dynamic recommendation scenarios as a sequence of temporal user-item interaction graphs. Formally, the dynamic graph is denoted by $\mathcal{G} = \{G_t\}_{t=1}^T$, where each snapshot at time step $t$ is represented as $G_t = (\mathcal{V}_t, \mathcal{E}_t, \mathcal{X}_t, \mathcal{A}_t)$. Here, $\mathcal{V}_t$ denotes the node set, $\mathcal{E}_t$ the edge set, $\mathcal{X}_t$ the feature matrix, and $\mathcal{A}_t$ the adjacency matrix at time $t$. To facilitate temporal modeling and evaluation, the complete temporal graph $\mathcal{G}$ is partitioned along the time axis into a training set $\mathcal{G}_{\text{train}}$ and a test set $\mathcal{G}_{\text{test}}$~\cite{jiang2024ragraph}.
Given a temporal graph $\mathcal{G}$, we formulate dynamic graph recommendation as the task of learning a predictive model to forecast future user-item interactions. To improve generalization under temporal distribution shifts, we adopt an augmented learning framework in which the model enhances the query subgraph by retrieving and incorporating relevant subgraphs from historical interactions. 

\paragraph{Recommendation Subgraph Library.}
To improve structural generalization in dynamic recommendation, we construct a \textit{Recommendation Subgraph Library} from historical user-item interactions, utilizing FAISS-based embedding retrieval~\cite{douze2024faiss}. Each resource subgraph is extracted as a $k$-hop neighborhood centered around nodes involved in past interactions, forming a collection $\mathcal{G}_R = \{G(v_r)\}_{r=1}^{R}$. Specifically, each subgraph is defined as:
\begin{equation}
G(v_r) = \left(v_r, \mathcal{N}^{(k)}(v_r), \mathcal{X}_r, \mathcal{A}_r \right),
\end{equation}
where $v_r$ denotes the central node, and $\mathcal{N}^{(k)}(v_r)$ represents its $k$-hop neighborhood, including both user and item nodes. In our formulation, the embedding of each subgraph serves as the \emph{key}, while its graph structure and representation jointly constitute the \emph{value}.

\paragraph{Task Relevance Estimation.}
To retrieve semantically relevant subgraphs for a given recommendation query, we define a task relevance function that evaluates the utility of a candidate subgraph. 
Given a task $\mathcal{T}$, a query subgraph $G(v_q)$, and a candidate subgraph $G(v_r)$ from the resource pool, the task relevance score is defined as:
\begin{equation}
\textsc{Rel}(G(v_q), G(v_r) \mid \mathcal{T}) = \mathcal{R}_\theta\Bigl(G(v_q), G(v_r)\Bigl),
\end{equation}
where $\mathcal{R}_\theta$ is a task-aware neural model parameterized by $\theta$, designed to jointly encode both the query graph and the candidate subgraph as dual inputs.

\section{TarDGR Framework}

\begin{figure}[t]
    \centering
    \includegraphics[width=1\linewidth]{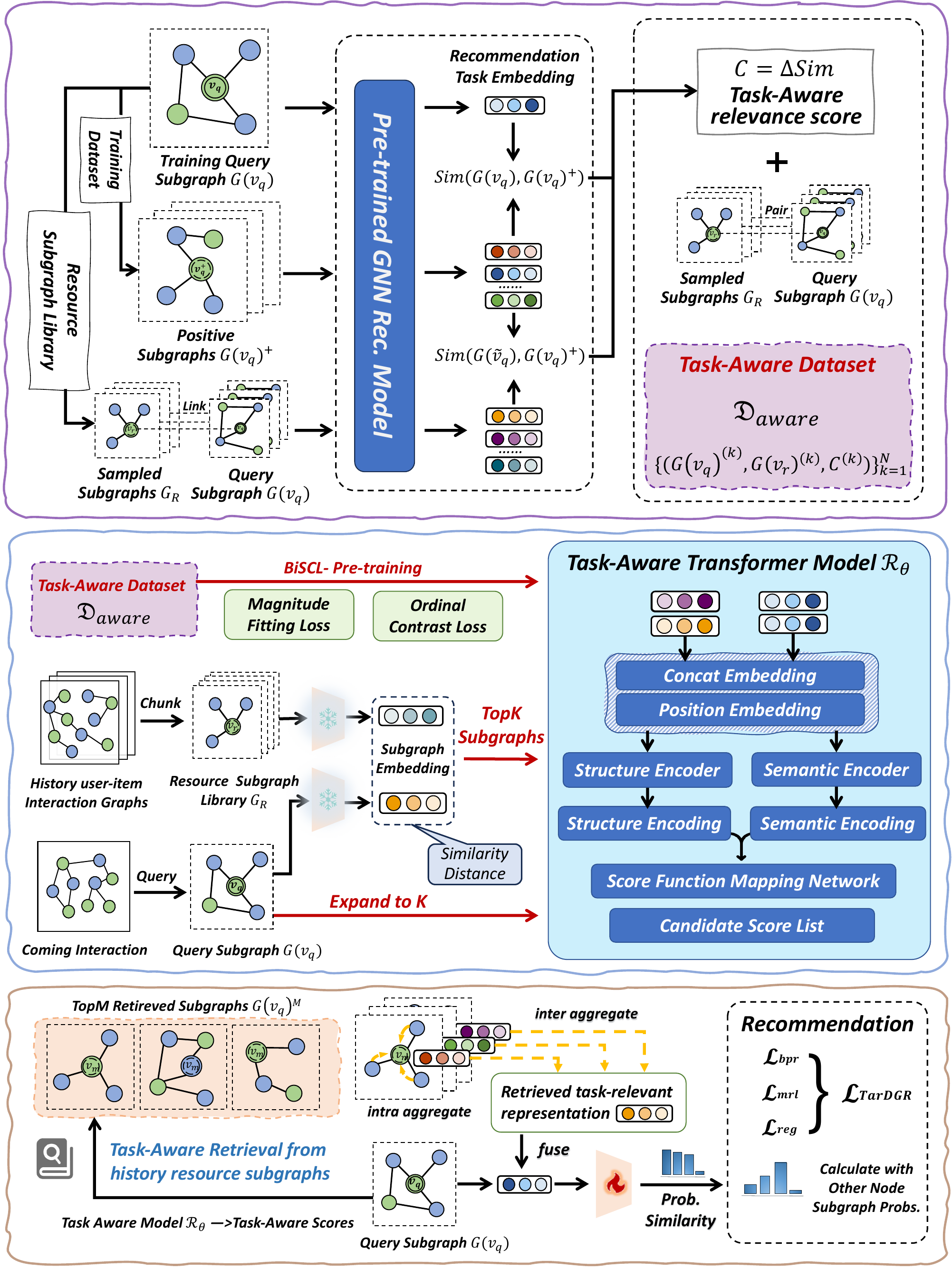}
    \caption{Overview of the {TarDGR} framework.}
    \label{fig:overall_TarDGR}
    \vspace{-0.2cm}
\end{figure}

In this section, we present \textbf{TarDGR}, a novel task-aware retrieval framework designed to enhance the generalization ability of dynamic graph recommendations. As illustrated in Figure~\ref{fig:overall_TarDGR}, the top depicts \textit{the {evaluation mechanism} part}, which automatically constructs task-specific supervision signals. The middle presents \textit{the {model} part}, detailing the input formulation, architectural design, and BiSCL pretraining. The bottom presents \textit{the {inference and training} pipeline}, where task-aware subgraphs are retrieved and integrated to enhance dynamic recommendation.
A formal theoretical analysis of the generalization benefits of TarDGR is provided in Appendix~\ref{appendix:theory_analysis}.

\subsection{Task-Aware Evaluation Mechanism}

We propose an automated task-aware evaluation mechanism that quantifies the contribution of a candidate historical subgraph $G(v_r)$ to the current recommendation query $G(v_q)$. 

Using a pre-trained GNN recommendation model $f^{\theta}_{\text{pre}}$, both subgraphs are encoded into embeddings $\textsc{Enc}\bigl(G(v_q)\bigl)$ and $\textsc{Enc}\bigl(G(v_r)\bigl)$. We first compute the average cosine similarity between the query embedding and a set of positive historical subgraphs $\{ G(v_q)^+ \}$:
\begin{equation}
\small
\overline{\textsc{Sim}}_{\text{before}} = \frac{1}{N^+} \sum_{i=1}^{N^+} \textsc{Cos}\Bigl(\textsc{Enc}\bigl(G(v_q)\bigl), \textsc{Enc}\bigl(G(v_q)^+_i\bigl)\Bigr).
\end{equation}

Next, we fuse the query subgraph $G(v_q)$ with the candidate subgraph $G(v_r)$ by constructing inter-subgraph links between the central modes and applying graph convolution to obtain a combined representation:
\begin{equation}
\textsc{Enc}\bigl(\tilde{G(v_q)}\bigl) = f_{\text{fuse}}\Bigl(\textsc{Enc}\bigl(G(v_q)\oplus G(v_r)\bigl)\Bigl).
\end{equation}

The updated similarity to the positive set is then:
\begin{equation}
\small
\overline{\textsc{Sim}}_{\text{after}} = \frac{1}{N^+} \sum_{i=1}^{N^+} \textsc{Cos}\bigl(\textsc{Enc}\bigl(\tilde{G(v_q)}\bigl), \textsc{Enc}\bigl((G(v_q)^+_i\bigl)\bigr).
\end{equation}
We define the relative similarity shift as: $\Delta \textsc{Rel} = \overline{\textsc{Sim}}_{\text{after}} - \overline{\textsc{Sim}}_{\text{before}}$. This $\Delta \textsc{Rel}$ is utilized as the task relevance score $C_r$, quantifying the degree to which the candidate subgraph $G(v_r)$ contributes to current recommendation. Specifically:
\begin{itemize}
    \item If $\Delta \textsc{Rel} > 0$, the candidate subgraph is positively correlated with the task and is considered a {beneficial} sample.
    \item If $\Delta \textsc{Rel} \approx 0$, it is deemed {irrelevant}.
    \item If $\Delta \textsc{Rel} < 0$, it is {negatively correlated} and considered {harmful} to task performance.
\end{itemize}
Based on this scoring, we construct a task-aware dataset $\mathcal{D}_{\text{aware}} = \{(G(v_q), G(v_r), C_r)\}$, where each triplet consists of a query subgraph, a candidate recommendation subgraph, and their associated task relevance score. This dataset provides task-aligned supervision signals for the subsequent training of a task-aware graph recommendation model.

\subsection{Graph Transformer-based Task-Aware Model}
\label{GTransTAM}
To equip the dynamic graph recommendation system with task-aware retrieval augmentation, we propose the \textbf{Graph Transformer-based Task-Aware Model} denoted as $\mathcal{R}_\theta$. This model is designed to effectively capture the complex interactions between a query subgraph and candidate subgraphs by leveraging the expressive power of graph transformers, thereby enhancing the relevance estimation and retrieval performance in dynamic recommendation scenarios.

\subsubsection{Subgraph Semantic Encoder}
To capture the temporal evolution of user preferences and interaction patterns, we initialize node embeddings using a pre-trained dynamic GNN~\cite{yang2024graphpro}, which encodes historical temporal dependencies across graph snapshots. Specifically, the initial node representations at the fine-tuning step $t$ are obtained via forward propagation over the graph at time $t-1$: ${h}_t = \text{forward}({h}_{t-1}; G_{t-1})$.

Each resource subgraph $G(v_r)$ is encoded by applying $L$-layer graph convolutions on its temporally contextualized node embeddings to yield subgraph-level representations:
\begin{equation}\label{subgraph_rep} 
h_r = \sum_{l=0}^L \textsc{GConv}(h_t^r, \mathcal{G}(v_r)) \in \mathbb{R}^d.
\end{equation}

Given the query subgraph embedding $h_q \in \mathbb{R}^d$, we compute pairwise L2 distances to resource subgraph embeddings $\{ h_r \}_{r=1}^R$: $\mathrm{dist}({h}_q, {h}_r) = {h}_q^\top {h}_q + {h}_r^\top {h}_r - 2 {h}_q^\top {h}_r$. The top-$K$ resource subgraphs with the smallest distances to the query subgraph are retrieved to form the initial candidate set:
\begin{equation}
\mathcal{G}(v_q)^K = \textsc{TopK}_{\text{search}} \left( {G}(v_q), {G}(v_r) \right).
\end{equation}

Each candidate is paired with the query subgraph to form the matching set: $\left\{ \left\langle {G}(v_q), {G}(v_i) \right\rangle \,|\, {G}(v_i) \in \mathcal{G}(v_q)^K \right\}$. To enable pairwise semantic modeling, we jointly encode both the query and candidate subgraphs:
\begin{equation}
\small
{h} = \left[ \sum_{l=0}^L \textsc{GConv}(h_t^q, \mathcal{G}(v_q)) \,\middle\|\, \sum_{l=0}^L \textsc{GConv}(h_t^i, \mathcal{G}(v_i)) \right].
\end{equation}

A positional embedding ${P}$ is added to ${h}$ to encode the relative positional order, ${h}_{\text{pos}} = {h} + {P}$. We input ${h}_{\text{pos}}$ into a multi-head self-attention module to capture fine-grained relational dependencies between query and candidate subgraphs. It is projected into query, key, and value matrices: ${Q}={h}_{\text{pos}}{W}_Q, {K}={h}_{\text{pos}}{W}_K,{V}={h}_{\text{pos}}{W}_V$. Attention weights are computed as: $\small\text{Attn} = \textsc{Softmax}\left(\frac{{Q} {K}^\top}{\sqrt{d_k}}\right){V}$, where $d_k$ represents the dimensionality of the key vectors. The final subgraph-level semantic representation is:
\begin{equation}
\label{eq:sem_concat}
{h}_{\text{sem}} = \textsc{Concat}(\text{Attn}_1, \dots, \text{Attn}_H) {W}_O,
\end{equation}
where ${W}_O$ is a trainable output projection matrix. The resulting embedding ${h}_{\text{sem}}$ serves as the task-aware semantic encoding of the query-candidate subgraph pair.

\subsubsection{Subgraph Structure Encoder}
We further employ a dedicated {subgraph structure encoder} to encode structural dependencies within each subgraph. The positional-enhanced embedding ${h}_{\text{pos}} \in \mathbb{R}^{2d}$ is first linearly projected into a lower-dimensional latent space: 
${h}_{\text{hid}} = {h}_{\text{pos}} {W} + {b}$. 

We then apply multi-layer, multi-head attention to capture fine-grained dependencies. At the $l$-th layer, query, key, and value matrices are computed via linear transformations, followed by attention calculation and concatenation across heads. Each attention layer is followed by residual connection and layer normalization: 
\begin{equation}
    {h}_{\text{hid}}^{(l+1)} = \textsc{LayerNorm}({h}_{\text{hid}}^{(l)} + \text{Attn}^{(l)}_{\text{output}})
\end{equation}

We apply position-wise feedforward network (FFN) to enhance representation and introduce non-linearity. The final FFN output is denoted as ${h}_{\text{ffn}}$. To encode subgraph-level structural patterns, the FFN output is aggregated via normalized adjacency propagation:
\begin{equation}
\label{eq:str_adj_pro}
{h}_{\text{str}} = \mathcal{D}^{-1} (\mathcal{A}_{\text{s}} + \mathbf{I}) {h}_{\text{ffn}} {W},
\end{equation}
where ${\mathcal{D}}$ is the degree matrix, $\mathcal{A}_{\text{s}}$ is corresponding adjacency matrix and ${W} \in \mathbb{R}^{d_{\text{hid}} \times d}$. 

The semantic encoding ${h}_{\text{sem}}$ and structural encoding ${h}_{\text{str}}$ are concatenated to form a task-aware fused representation: ${h}_{\text{task}} = \textsc{Concat}({h}_{\text{sem}}, {h}_{\text{str}})$. The representation is transformed into a task-specific relevance score via a parametric scoring function $\mathcal{S}_\psi(\cdot)$:
\begin{equation}
s_i = \mathcal{S}_\psi({h}_{\text{task}}) = {w}^\top \textsc{ReLu}({W} {h}_{\text{task}} + {b}),
\end{equation}
where ${w}$ maps the hidden representation to a scalar score.

\subsubsection{BiSCL Pretraining of Task-Aware Model}
We introduce Bi-Level Supervised Correlation Loss (\textbf{BiSCL}), which pretrains the task-aware model by jointly supervising numerical fidelity and ordinal consistency. 

Given $\mathcal{D}_{\text{aware}} = {(G(v_q), G(v_r), C)}$, each subgraph is encoded into semantic-structural embeddings ${z}_q$ and ${z}_r$ as in Equation~\ref{subgraph_rep}. The concatenated pairwise feature ${h}_{q,r} = [{z}_q , | , {z}_r]$ and adjacency matrix $\mathcal{A}_{\text{s}}$ are then fed into $\mathcal{R}_\theta$ to compute the predicted task relevance: $\mathcal{R}_\theta({h}_{q,r}, \mathcal{A}_{\text{s}})$.

We apply a magnitude fitting loss to minimize the discrepancy between predicted and task relevance scores:
\begin{equation}
\mathcal{L}_{\text{mtl}} = \frac{1}{N} \sum_{k=1}^{N} \left( \mathcal{R}_\theta({h}_{q,r}, \mathcal{A}_{\text{s}}) - C \right)^2.
\end{equation}

In parallel, we impose a pairwise ordinal constraint loss to preserve inter-sample ordering. Specifically, for every pair $(k, l)$ such that $C^{(k)} > C^{(l)}$, the predicted 
scores must maintain this ranking:
\begin{equation}
\small
\mathcal{L}_{\text{ocl}} = \log \left[ 1 + \sum_{\substack{k, l }} \exp \left( \frac{\mathcal{R}_\theta({h}_{q,r}^{(l)}, \mathcal{A}_{\text{s}}) - \mathcal{R}_\theta({h}_{q,r}^{(k)}, \mathcal{A}_{\text{s}})}{\tau} \right) \right],
\end{equation}
where $\tau$ is a temperature hyperparameter controlling the penalty's smoothness. 
The BiSCL loss is expressed as:
\begin{equation}
\mathcal{L}_{\text{BiSCL}} = \rho \cdot \mathcal{L}_{\text{ocl}} + (1 - \rho) \cdot \mathcal{L}_{\text{mtl}} ,
\end{equation}
where $\rho \in [0, 1]$ balances absolute fidelity and ordinal coherence. BiSCL injects task-aware inductive signals, facilitating robust pretraining for downstream retrieval task.

\subsection{Task-Aware Retrieval Inference and Training}
Given a query node $v_q$ and subgraph $G(v_q)$, we obtain task-aware relevance score list $\{s^{(i)}\}_{i=1}^{K}$ by $\mathcal{R}_\theta$ based on task-aware model in Section~\ref{GTransTAM}. We select the top-$M$ subgraphs with the highest task-specific relevance scores to construct the set for retrieval augmentation $\left\{ {G}(v_m) \in \mathcal{G}(v_q)^M \right\}$. Through intra-graph aggregation, we obtain the internal representation of all task-relevant subgraphs: 
\begin{equation}
\label{eq:intra_agg}
h_m = \sum_{l=1}^{L} \textsc{GConv}^{(l)}(h_t, {G}(v_m)),
\end{equation}
where $h_t$ denotes features from dynamic encoder. Retrieved subgraphs are then aggregated via soft evidence aggregation:
\begin{equation}
\label{eq:soft_agg}
H_{\text{rag}} = \sum_{i=1}^{M} \alpha_i \cdot h_m^i, \quad \sum_{i=1}^{M} \alpha_i = 1,
\end{equation}
where $\alpha_i$ are normalized weights indicating retrieval confidence, either uniform or learned from $\mathcal{R}_\theta$ scores. Then, we employ a residual fusion mechanism to integrate retrieval representation into the query subgraph:
\begin{equation}
\label{eq:fuse_agg}
\tilde{h}_q = \beta h_q + (1 - \beta) H_{\text{rag}},
\end{equation}
where $\beta$ is a learnable gate balancing original and retrieved task-relevant representation.

Semantic retrieval of historical subgraphs recovers latent dependencies beyond the query 
context, improving generalization across temporal 
recommendation tasks. 
Finally, we conduct a joint fine-tuning recommendation loss function. Structural robustness is encouraged by injecting stochastic perturbations into the graph topology: $\mathcal{E}' = \{ (u, v) \in \mathcal{E} \mid \text{Bernoulli}(r) = 1 \}, \quad \mathcal{G}' = (\mathcal{V}, \mathcal{E}')$.

User preference is modeled using a task-aware scoring function trained under a margin-based objective. For each training triplet $(u, i^+, i^-)$, where $u$ and $i$ denote the user and item respectively, the corresponding subgraph inputs are constructed as: ${X}^+ = [{h}_u \, \| \, {h}_{i^+}]$ and ${X}^- = [{h}_u \, \| \, {h}_{i^-}]$, followed by relevance scoring via $\mathcal{R}_\theta$. The resulting margin ranking loss encourages the model to distinguish relevant items from irrelevant ones in a task-consistent manner:
\begin{equation}
\mathcal{L}_{\text{mrl}} = \frac{1}{|\mathcal{B}|} \sum_{(u, i^+, i^-) \in \mathcal{B}} \max(0, \gamma - (s^+ - s^-)).
\end{equation}

To mitigate overfitting and promote stable training, we apply a penalty to the embedding norms:
\begin{equation}
\small
\mathcal{L}_{\text{reg}} = \frac{1}{2N} \left( \sum_{u \in \mathcal{B}_u} \| {h}_u \|_2^2 + \sum_{i \in \mathcal{B}_i^+} \| {h}_i^+ \|_2^2 + \sum_{i \in \mathcal{B}_i^-} \| {h}_i^- \|_2^2 \right).
\end{equation}

The recommendation preference is captured via the bayesian personalized ranking loss~\cite{rendle2012bpr}:
\begin{equation}
\mathcal{L}_{\text{bpr}} = - \sum_{(u, i^+, i^-) \in \mathcal{B}} \log \sigma({h}_u^\top {h}_{i^+} - {h}_u^\top {h}_{i^-}),
\end{equation}
where $\sigma(\cdot)$ denotes the Sigmoid function. The final loss aggregates all components:
\begin{equation}
\mathcal{L}_{\text{total}} = \mathcal{L}_{\text{bpr}} + \lambda \cdot \mathcal{L}_{\text{mrl}} + \mu \cdot \mathcal{L}_{\text{reg}},
\end{equation}
where $\lambda$ and $\mu$ are weighting hyperparameters. This optimization encourages the model to align recommendation signals with task-aware semantics, while improving robustness under perturbation and preventing overfitting.

\begin{table*}[t]
    \centering
    \begin{tabular}{lcccccc}
        \toprule
        \multirow{2}{*}{Method} & \multicolumn{2}{c}{TAOBAO} & \multicolumn{2}{c}{KOUBEI} & \multicolumn{2}{c}{AMAZON} \\
        \cmidrule(lr){2-3}\cmidrule(lr){4-5}\cmidrule(lr){6-7}
        ~ & Recall & nDCG & Recall & nDCG & Recall & nDCG \\
        \midrule
        LightGCN & 22.47$\scriptstyle{\pm 02.53}$ & 21.89$\scriptstyle{\pm 02.80}$ & 30.21$\scriptstyle{\pm 06.45}$ & 22.24$\scriptstyle{\pm 05.83}$ & 15.07$\scriptstyle{\pm 06.48}$ & 06.53$\scriptstyle{\pm 02.66}$ \\ 
        SGL & 22.15$\scriptstyle{\pm 02.20}$ & 22.12$\scriptstyle{\pm 03.09}$ & 32.61$\scriptstyle{\pm 04.27}$ & 22.36$\scriptstyle{\pm 04.82}$ & 15.78$\scriptstyle{\pm 07.12}$ & 07.90$\scriptstyle{\pm 02.49}$ \\ 
        MixGCF & 22.84$\scriptstyle{\pm 02.15}$ & 23.05$\scriptstyle{\pm 03.87}$ & 32.06$\scriptstyle{\pm 04.20}$ & 22.49$\scriptstyle{\pm 06.91}$ & 15.24$\scriptstyle{\pm 08.98}$ & 07.40$\scriptstyle{\pm 03.44}$ \\ 
        SimGCL & 22.18$\scriptstyle{\pm 02.22}$ & 23.15$\scriptstyle{\pm 02.75}$ & 33.07$\scriptstyle{\pm 05.28}$ & 23.08$\scriptstyle{\pm 05.55}$ & 16.10$\scriptstyle{\pm 07.91}$ & 07.58$\scriptstyle{\pm 03.51}$ \\
        GraphPrompt  & 20.76$\scriptstyle{\pm 01.54}$ & 20.22$\scriptstyle{\pm 00.98}$ & {33.24}$\scriptstyle{\pm 04.98}$ & {24.12}$\scriptstyle{\pm 09.20}$ & {16.20}$\scriptstyle{\pm 08.58}$ & {07.89}$\scriptstyle{\pm 04.12}$ \\
        GPF          & 22.46$\scriptstyle{\pm 01.66}$ & 22.12$\scriptstyle{\pm 01.16}$ & {33.70}$\scriptstyle{\pm 06.44}$ & {24.39}$\scriptstyle{\pm 04.01}$ & {17.67}$\scriptstyle{\pm 09.04}$ & {08.94}$\scriptstyle{\pm 04.57}$ \\
        EvolveGCN-H  & 22.44$\scriptstyle{\pm 02.55}$ & 22.17$\scriptstyle{\pm 01.79}$ & {31.22}$\scriptstyle{\pm 04.25}$ & {23.00}$\scriptstyle{\pm 02.92}$ & {14.97}$\scriptstyle{\pm 10.28}$ & {07.20}$\scriptstyle{\pm 05.43}$ \\
        EvolveGCN-O  & 23.64$\scriptstyle{\pm 02.13}$ & 23.24$\scriptstyle{\pm 01.28}$ & {33.01}$\scriptstyle{\pm 05.22}$ & {23.98}$\scriptstyle{\pm 04.01}$ & {17.48}$\scriptstyle{\pm 08.13}$ & {08.68}$\scriptstyle{\pm 04.25}$ \\
        ROLAND       & 22.67$\scriptstyle{\pm 02.42}$ & 22.60$\scriptstyle{\pm 01.91}$ & {30.11}$\scriptstyle{\pm 03.14}$ & {22.29}$\scriptstyle{\pm 01.84}$ & {15.33}$\scriptstyle{\pm 07.10}$ & {07.09}$\scriptstyle{\pm 03.02}$ \\
        \cmidrule(lr){1-7}
        \textbf{GraphPro}+ \\
        \cmidrule(lr){1-7}
        Vanilla/NF  & 20.10$\scriptstyle{\pm 01.50}$ & 20.12$\scriptstyle{\pm 01.30}$ & 21.31$\scriptstyle{\pm 04.59}$ & 15.31$\scriptstyle{\pm 03.11}$ & 12.56$\scriptstyle{\pm 07.45}$ & 06.31$\scriptstyle{\pm 03.92}$ \\
        Vanilla/FT  & {23.99}$\scriptstyle{\pm 02.11}$ & 23.26$\scriptstyle{\pm 01.42}$ & {33.96}$\scriptstyle{\pm 04.13}$ & {24.66}$\scriptstyle{\pm 02.78}$ & 18.14$\scriptstyle{\pm 07.55}$ & 08.73$\scriptstyle{\pm 03.74}$ \\
        PRODIGY/NF  & 21.67$\scriptstyle{\pm 01.42}$ & 23.15$\scriptstyle{\pm 03.20}$ & 21.66$\scriptstyle{\pm 03.21}$ & 14.82$\scriptstyle{\pm 03.92}$ & 11.88$\scriptstyle{\pm 02.61}$ & 05.84$\scriptstyle{\pm 01.84}$ \\
        PRODIGY/FT  & {23.74}$\scriptstyle{\pm 01.22}$ & 23.65$\scriptstyle{\pm 02.31}$ &{33.46}$\scriptstyle{\pm 04.70}$ &{23.28}$\scriptstyle{\pm 03.40}$ & 16.72$\scriptstyle{\pm 04.28}$ &{08.09}$\scriptstyle{\pm 02.66}$ \\
        RAGRAPH/NF  & 20.31$\scriptstyle{\pm 01.60}$ & 20.45$\scriptstyle{\pm 01.44}$ & 22.86$\scriptstyle{\pm 03.44}$ & 16.68$\scriptstyle{\pm 02.48}$ & 13.78$\scriptstyle{\pm 05.54}$ & 06.52$\scriptstyle{\pm 02.69}$ \\
        RAGRAPH/FT  &\underline{24.78}$\scriptstyle{\pm 01.93}$ &\underline{24.35}$\scriptstyle{\pm 01.34}$ & \underline{34.27}$\scriptstyle{\pm 03.93}$ & \underline{24.82}$\scriptstyle{\pm 02.69}$ &\underline{18.69}$\scriptstyle{\pm 07.45}$ & \underline{09.09}$\scriptstyle{\pm 03.89}$ \\
        \cmidrule(lr){1-7}
        TarDGR/NF    & 20.39$\scriptstyle{\pm 02.41}$ & 20.91$\scriptstyle{\pm 02.18}$ & {24.83}$\scriptstyle{\pm 03.68}$ & {17.90}$\scriptstyle{\pm 02.62}$ & {14.26}$\scriptstyle{\pm 05.37}$ & {06.74}$\scriptstyle{\pm 02.56}$ \\
        TarDGR/FT    & \textbf{25.20}$\scriptstyle{\pm 02.13}$ & \textbf{24.59}$\scriptstyle{\pm 01.42}$ & \textbf{36.52}$\scriptstyle{\pm 04.44}$ & \textbf{26.63}$\scriptstyle{\pm 02.98}$ & \textbf{19.56}$\scriptstyle{\pm 07.17}$ & \textbf{09.70}$\scriptstyle{\pm 03.62}$ \\
        \bottomrule
    \end{tabular}
    \caption{Main performance comparison results of TarDGR. The best performance is bolded, and the second is underlined.}
    \label{tab:overall}
    \vspace{-0.2cm}
\end{table*}

\section{Experiments}
\label{sec:experiments}
This section presents experiments designed to evaluate the performance of \textbf{TarDGR}, against state-of-the-art baselines on three dynamic graph datasets. Further experiments and analyses are provided in the Appendix.

\subsection{Experimental Setup}
\label{sec:exp setup}
\noindent\textbf{Datasets.} 
We evaluate our method on three public datasets spanning diverse dynamic recommendation scenarios: \textit{Taobao}, with 10 days of implicit feedback from the Taobao platform; \textit{Koubei}, a 9-week user–store interaction dataset from Alipay’s location service released for IJCAI’16; and \textit{Amazon}, containing 13 weeks of product review data. Additional details are provided in Appendix~\ref{appendix: datasets}.

\noindent\textbf{Methods and Baselines.}
We compare our approach against representative baselines spanning four major categories:
GNN-based recommenders, including LightGCN~\cite{he2020lightgcn} and its self-supervised variants such as SGL~\cite{sgl}, MixGCF~\cite{huang2021mixgcf}, and SimGCL~\cite{simgcl};
Dynamic GNNs, including EvolveGCN-H/O~\cite{pareja2020evolvegcn}, ROLAND~\cite{you2022roland}, and GraphPro~\cite{yang2024graphpro};
Graph prompting models, including GraphPrompt~\cite{graphprompt} and GPF~\cite{fang2023universal}; and
Retrieval-augmented models, such as PRODIGY~\cite{mishchenko2023prodigy} and RAGRAPH~\cite{jiang2024ragraph}, where retrieved subgraphs are integrated into GraphPro to enhance contextual representation.
Further details on the baselines are provided in Appendix~\ref{appendix:baselines}.

\noindent\textbf{Settings and Evaluation.}
We adopt the pre-trained dynamic graph dataset as the resource pool. For retrieval-based methods, we consider two variants: non-fine-tuned (NF), which applies plug-and-play retrieval augmentation without additional training on the target dataset, and fine-tuned (FT), which applies tuning on the training set. All models are pre-trained on historical snapshots and subsequently fine-tuned and evaluated on future snapshots. We report average performance over time using Recall@20 and nDCG@20~\cite{he2020lightgcn, simgcl}.
Further evaluation metrics and experimental settings are detailed in 
Appendix~\ref{appendix:evaluation-metrics},~\ref{appendix:setting-parameter}.

\begin{figure}[ht]
    \vspace{-0.3cm}
    \centering
    \subfigure[Recall]{
    \begin{minipage}[t]{0.49\linewidth}
    \centering
    \includegraphics[width=\linewidth]{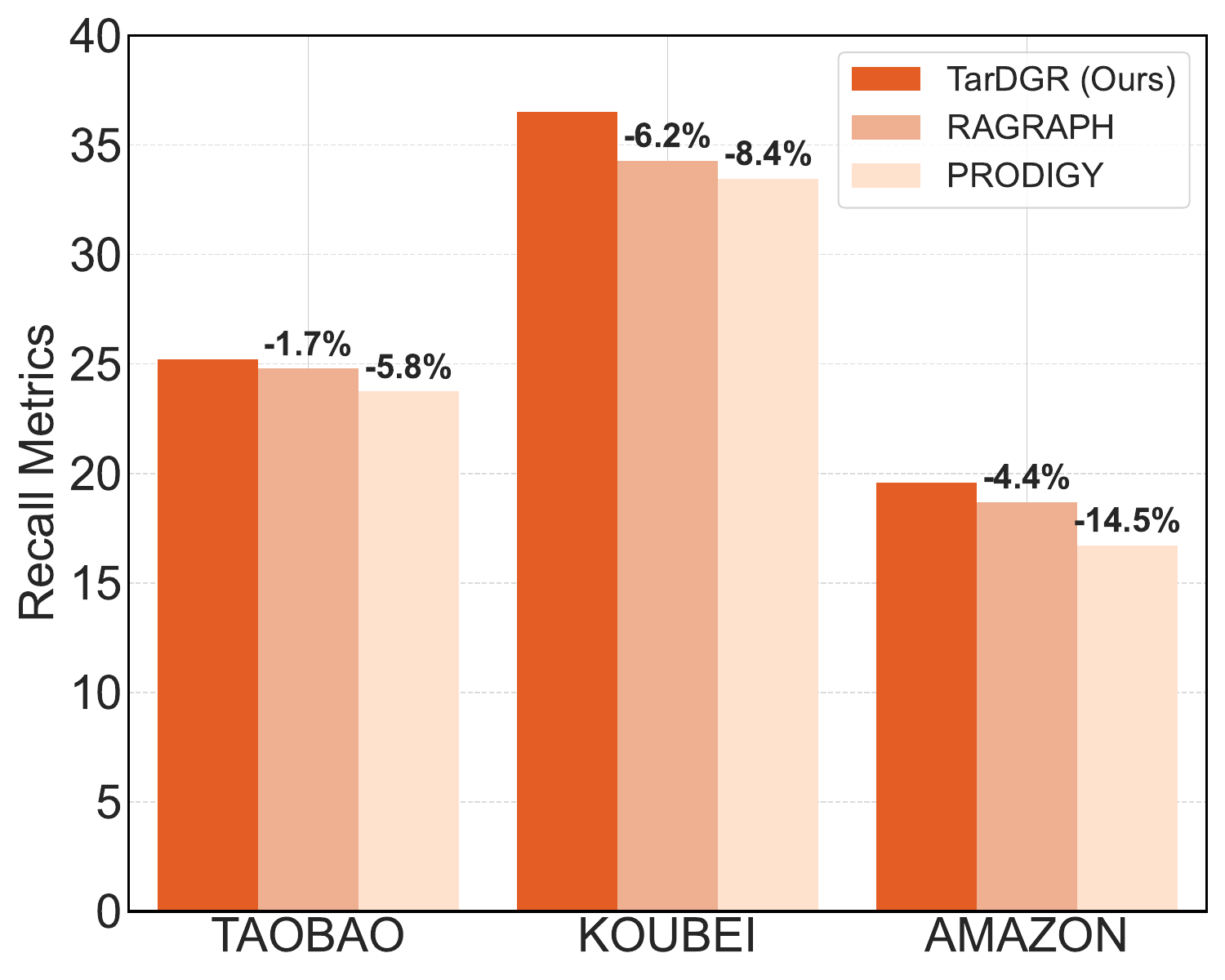}
    \end{minipage}%
    }%
    \subfigure[nDCG]{
    \begin{minipage}[t]{0.49\linewidth}
    \centering
    \includegraphics[width=\linewidth]{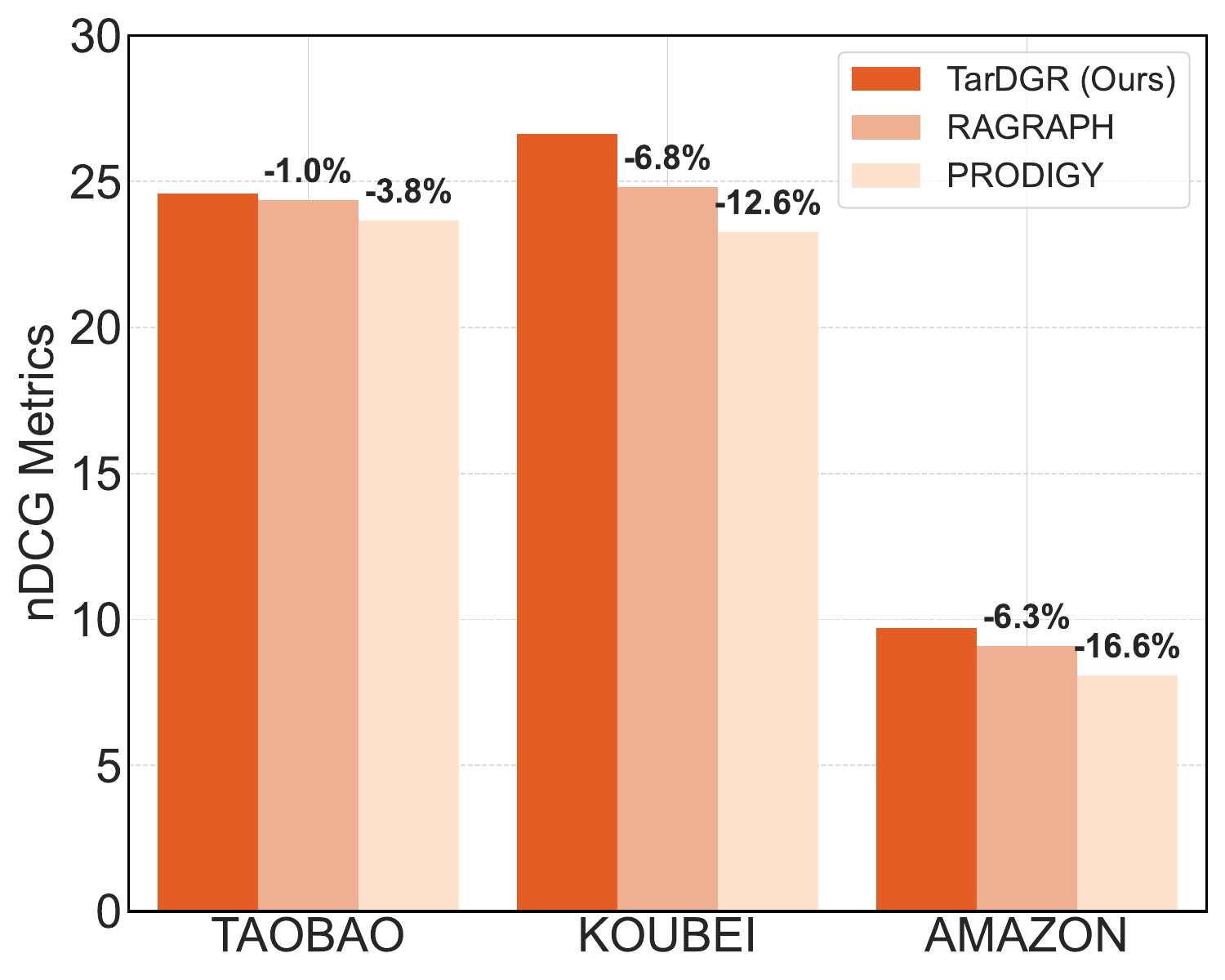}
    \end{minipage}
    }%
    \centering
\caption{Performance comparison of TarDGR and other RAG methods.}
\label{tab:RAG_method}
\vspace{-0.7cm}
\end{figure}

\subsection{Baseline Performance Comparison}
\label{sec:expt:perf}

\begin{figure}[ht]
    \vspace{-0.1cm}
    \centering
    \subfigure[Recall]{
    \begin{minipage}[t]{0.49\linewidth}
    \centering
    \includegraphics[width=\linewidth]{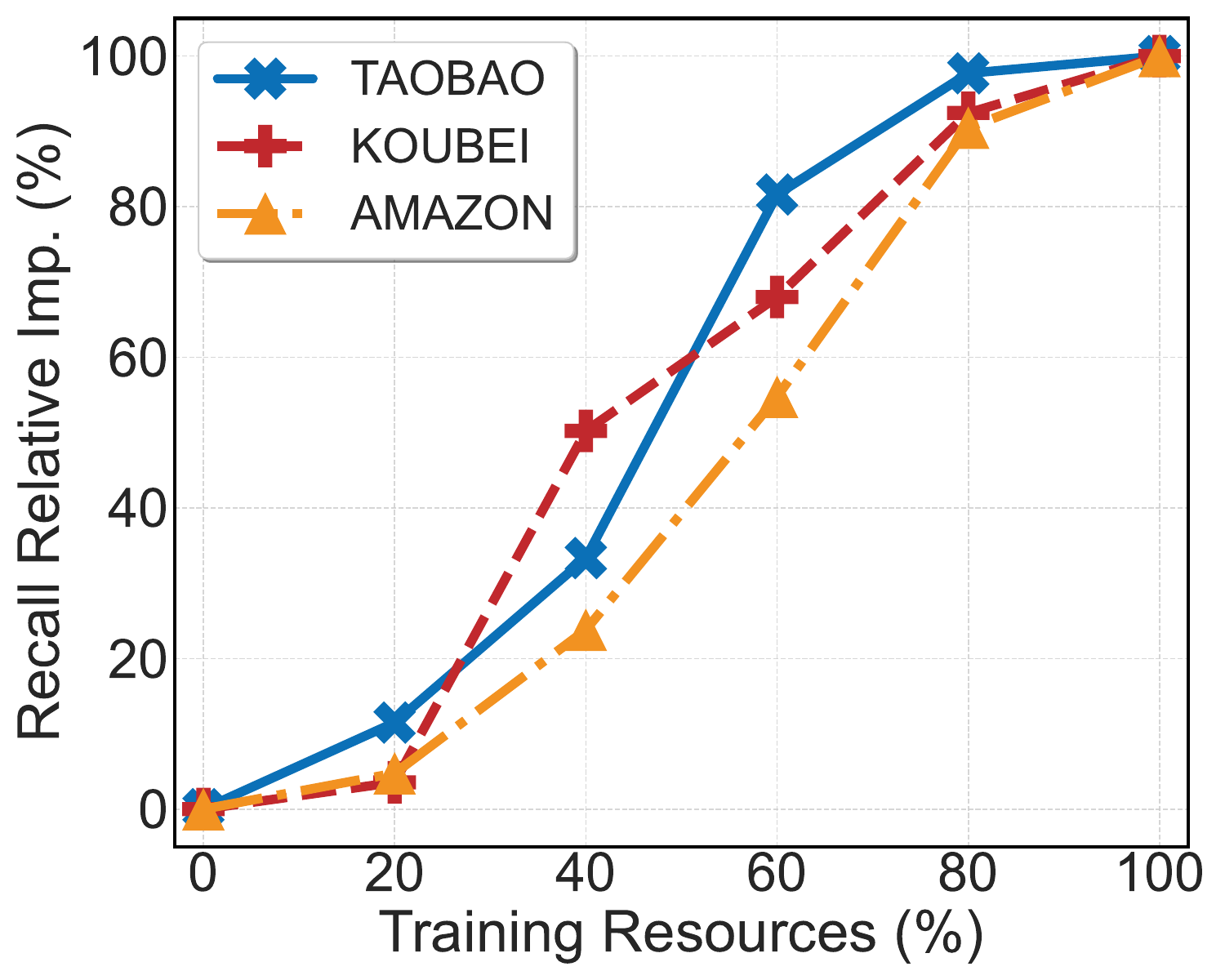}
    \end{minipage}%
    }%
    \subfigure[nDCG]{
    \begin{minipage}[t]{0.49\linewidth}
    \centering
    \includegraphics[width=\linewidth]{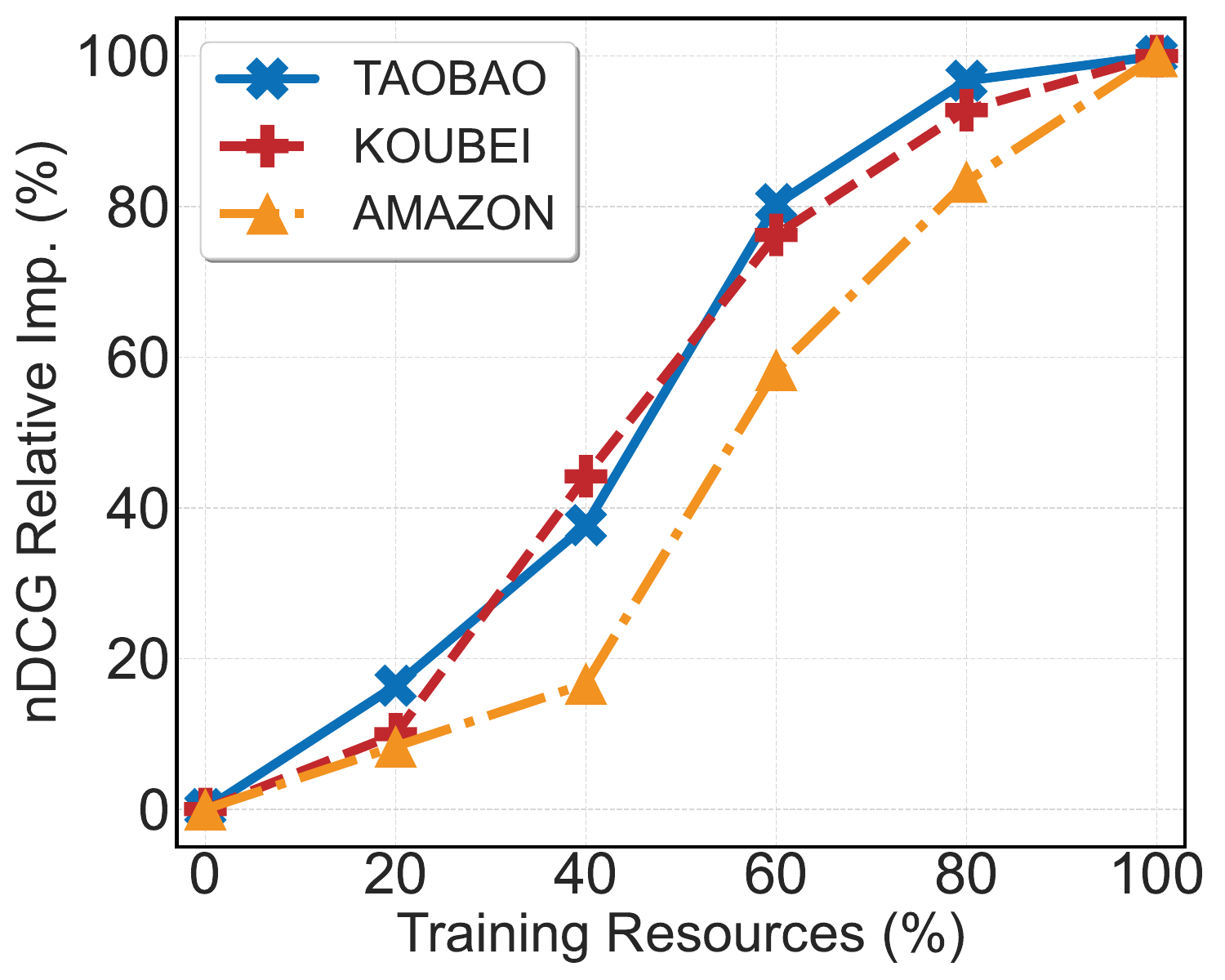}
    \end{minipage}
    }%
    \centering
\caption{Training resource experiments for the Graph Transformer-based Aware Model applied to TarDGR.}
\label{fig:train_res}
\vspace{-0.5cm}
\end{figure}

Table~\ref{tab:overall} summarizes the main experimental results, where TarDGR consistently outperforms all baselines across the three benchmark datasets. TarDGR/FT achieves the best overall performance, verifying the efficacy of our task-aware retrieval mechanism. By leveraging semantically aligned subgraphs from prior temporal resources, the model effectively transfers history knowledge to downstream recommendation tasks, reinforcing the value of task-aware enhancement in dynamic recommendation settings. 
By incorporating task-aware retrieval into the GraphPro framework, TarDGR achieves significant gains over RAGRAPH and PRODIGY. On Amazon, as shown in Table~\ref{tab:RAG_method}, it outperforms PRODIGY by 16.6\% in nDCG and 14.5\% in Recall, and RAGRAPH by 6.3\% and 4.4\%, respectively. These improvements highlight the importance of injecting task-specific subgraphs to enhance temporal generalization.

\begin{table*}[ht]
\centering
\begin{tabular}{lcccccc}
        \toprule
        \multirow{2}{*}{Method} & \multicolumn{2}{c}{TAOBAO} & \multicolumn{2}{c}{KOUBEI} & \multicolumn{2}{c}{AMAZON} \\
        \cmidrule(lr){2-3}\cmidrule(lr){4-5}\cmidrule(lr){6-7}
        ~ & Recall & nDCG & Recall & nDCG & Recall & nDCG \\
        \midrule
        w/o all & 24.63$\scriptstyle{\pm 01.81}$ & 24.02$\scriptstyle{\pm 01.79}$ & 34.14$\scriptstyle{\pm 03.57}$ & 24.83$\scriptstyle{\pm 02.13}$ & 18.42$\scriptstyle{\pm 06.71}$ & 08.91$\scriptstyle{\pm 03.04}$ \\
        w/o SEM & 24.95$\scriptstyle{\pm 01.75}$ & 24.48$\scriptstyle{\pm 01.57}$ & 35.84$\scriptstyle{\pm 03.88}$ & 25.56$\scriptstyle{\pm 02.53}$ & 19.10$\scriptstyle{\pm 07.52}$ & 09.45$\scriptstyle{\pm 03.59}$ \\
        w/o STR & 25.14$\scriptstyle{\pm 02.10}$ & 24.52$\scriptstyle{\pm 01.63}$ & 36.30$\scriptstyle{\pm 03.61}$ & 26.12$\scriptstyle{\pm 02.75}$ & 19.42$\scriptstyle{\pm 06.88}$ & 09.57$\scriptstyle{\pm 03.95}$ \\
        TarDGR & 25.20$\scriptstyle{\pm 02.13}$ & 24.59$\scriptstyle{\pm 01.42}$ & 36.52$\scriptstyle{\pm 04.44}$ & 26.63$\scriptstyle{\pm 02.98}$ & 19.56$\scriptstyle{\pm 07.17}$ & 09.70$\scriptstyle{\pm 03.62}$ \\
        \bottomrule
\end{tabular}
\caption{Ablation Study on Graph Transformer-based Aware Model.}
\label{tab:ab_study}
\vspace{-0.3cm}
\end{table*}

\begin{figure}[htbp]
\vspace{-0.2cm}
\centering
\subfigure[Taobao]{
\begin{minipage}[t]{0.49\linewidth}
\centering
\includegraphics[width=\linewidth]{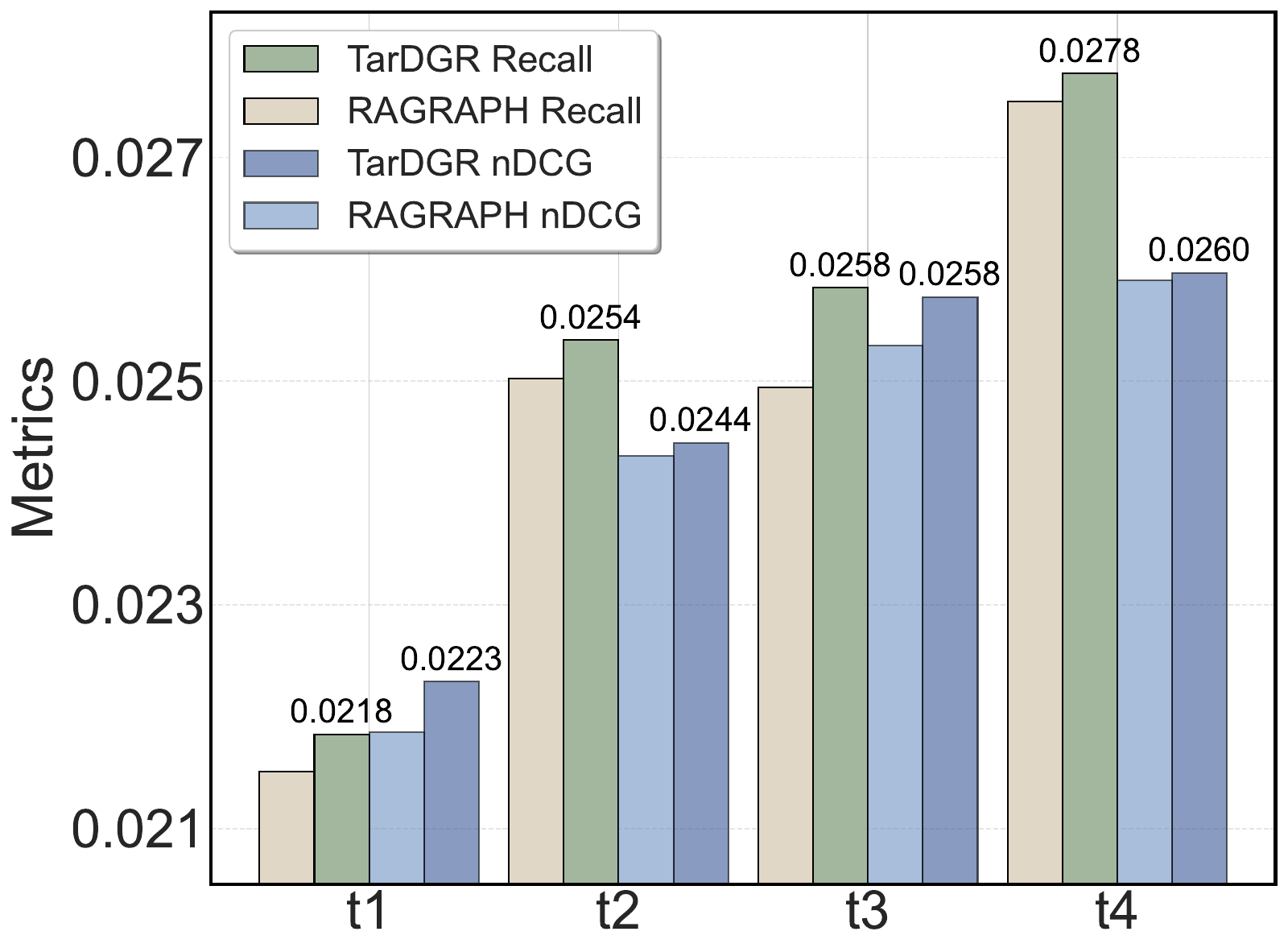}
\end{minipage}%
}%
\subfigure[Koubei]{
\begin{minipage}[t]{0.49\linewidth}
\centering
\includegraphics[width=\linewidth]{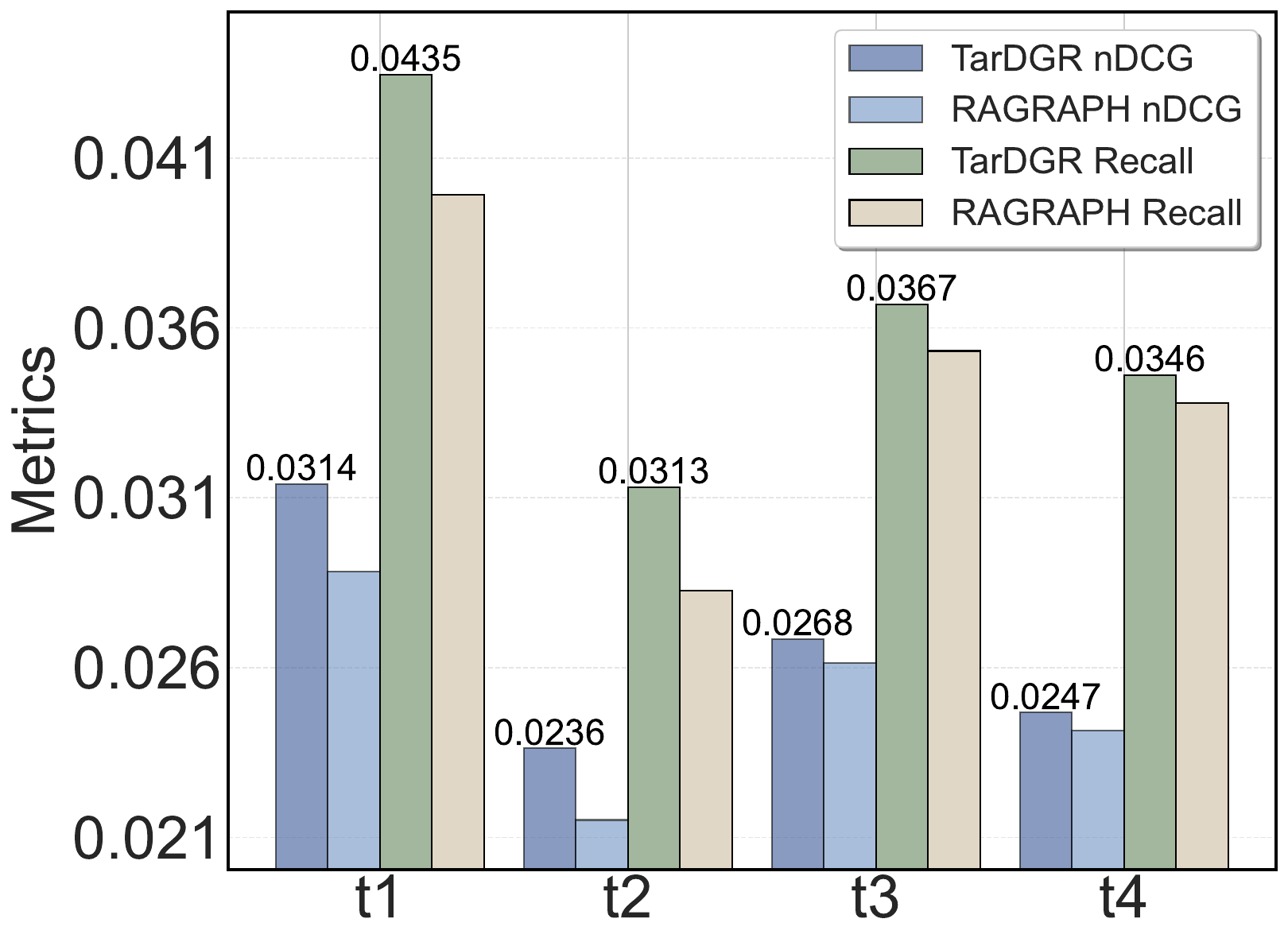}
\end{minipage}
}%

\subfigure[Amazon]{
\begin{minipage}[t]{0.99\linewidth}
\centering
\includegraphics[width=\linewidth]{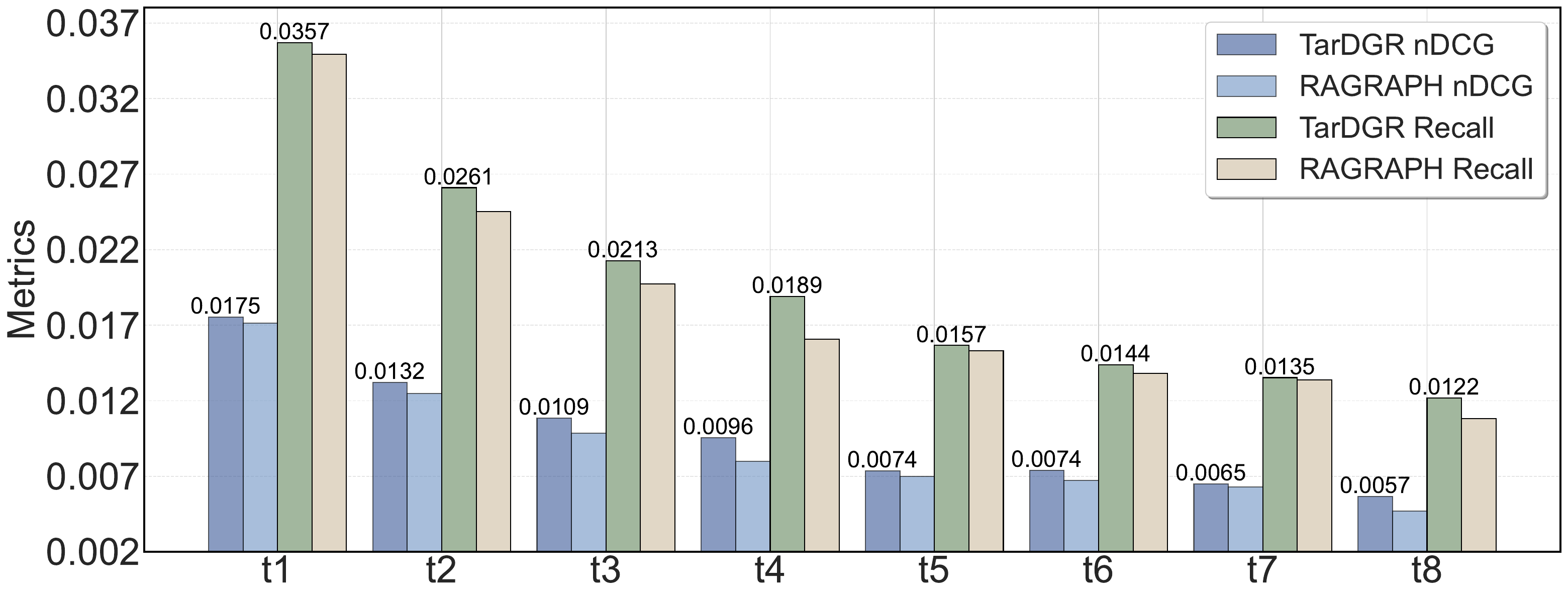}
\end{minipage}
}%
\centering
\caption{Comparison of TarDGR/FT and RAGRAPH/FT performance on each time step.}
\label{fig:each_step}
\vspace{-0.4cm}
\end{figure}

To further validate generalization over time, we provide a detailed comparison across individual time steps in Figure~\ref{fig:each_step} between TarDGR and RAGRAPH. TarDGR consistently demonstrates stronger performance across all time snapshots, particularly during earlier stages where user interactions are more closely aligned with resource subgraphs. This indicates that TarDGR benefits from enhanced transferability of task-relevant signals from historical contexts. 

\subsection{Ablation Study of TarDGR}

\noindent\textbf{Impact of Semantic and Structural Encoding.} 
Table~\ref{tab:ab_study} presents an ablation study assessing the contributions of the semantic and structural encoders within the TarDGR framework. 
The removal of both components ({w/o all}) causes a significant performance drop, affirming the necessity of task-aware representation learning. 
Removing the semantic encoder ({w/o SEM}) results in a larger performance drop than removing the structural encoder ({w/o STR}), indicating that capturing semantic relevance between subgraph embeddings via attention is particularly vital for generating accurate relevance scores. Nonetheless, omitting structural encoding ({w/o STR}) also causes noticeable degradation, demonstrating its complementary value in modeling subgraph-level structural compatibility.

\noindent\textbf{Effect of Task-Aware Resource Weighting.}  
We examine the impact of task-aware training resource weight in TarDGR, as shown in Figure~\ref{fig:train_res}. The performance follows a non-linear trend: initial improvements are slow due to limited supervision; as more subgraphs are introduced, the model quickly gains expressiveness and improves; finally, performance saturates when the resource pool becomes redundant or overly dense. The trend underscores the importance of task-aligned supervision and highlights that the quantity of retrieved training subgraphs are essential for maximizing the effectiveness of the TarDGR framework. This trend underscores the importance of task-aligned supervision and validates our retrieval-based training strategy for enhancing generalization in dynamic recommendation.

\begin{figure}[ht]
\vspace{-0.3cm}
    \centering
    \subfigure[Hyper-parameter Top-$M$]{
    \begin{minipage}[t]{0.49\linewidth}
    \centering
    \includegraphics[width=\linewidth]{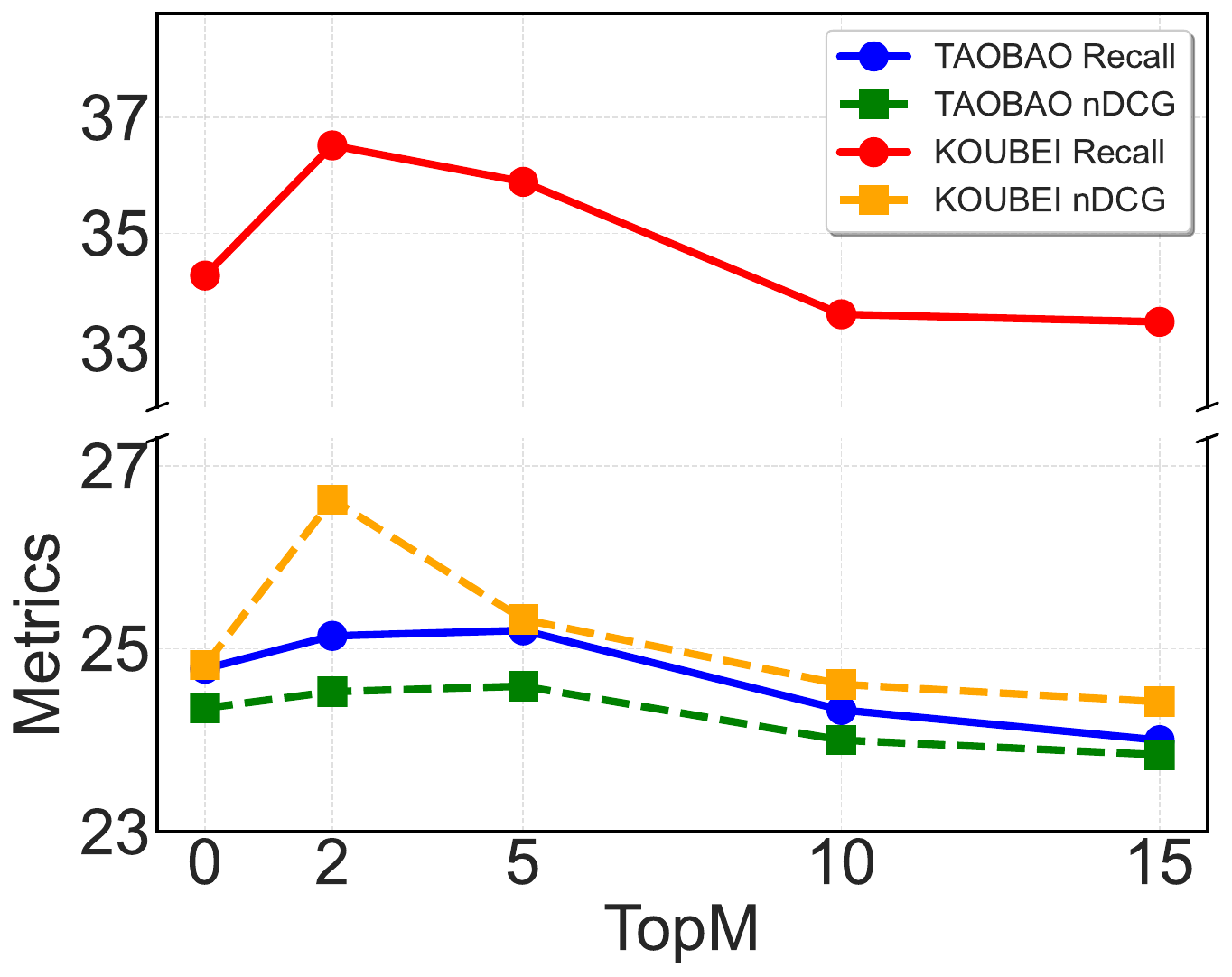}
    \end{minipage}%
    }%
    \subfigure[Hyper-parameter $k$-hop]{
    \begin{minipage}[t]{0.49\linewidth}
    \centering
    \includegraphics[width=\linewidth]{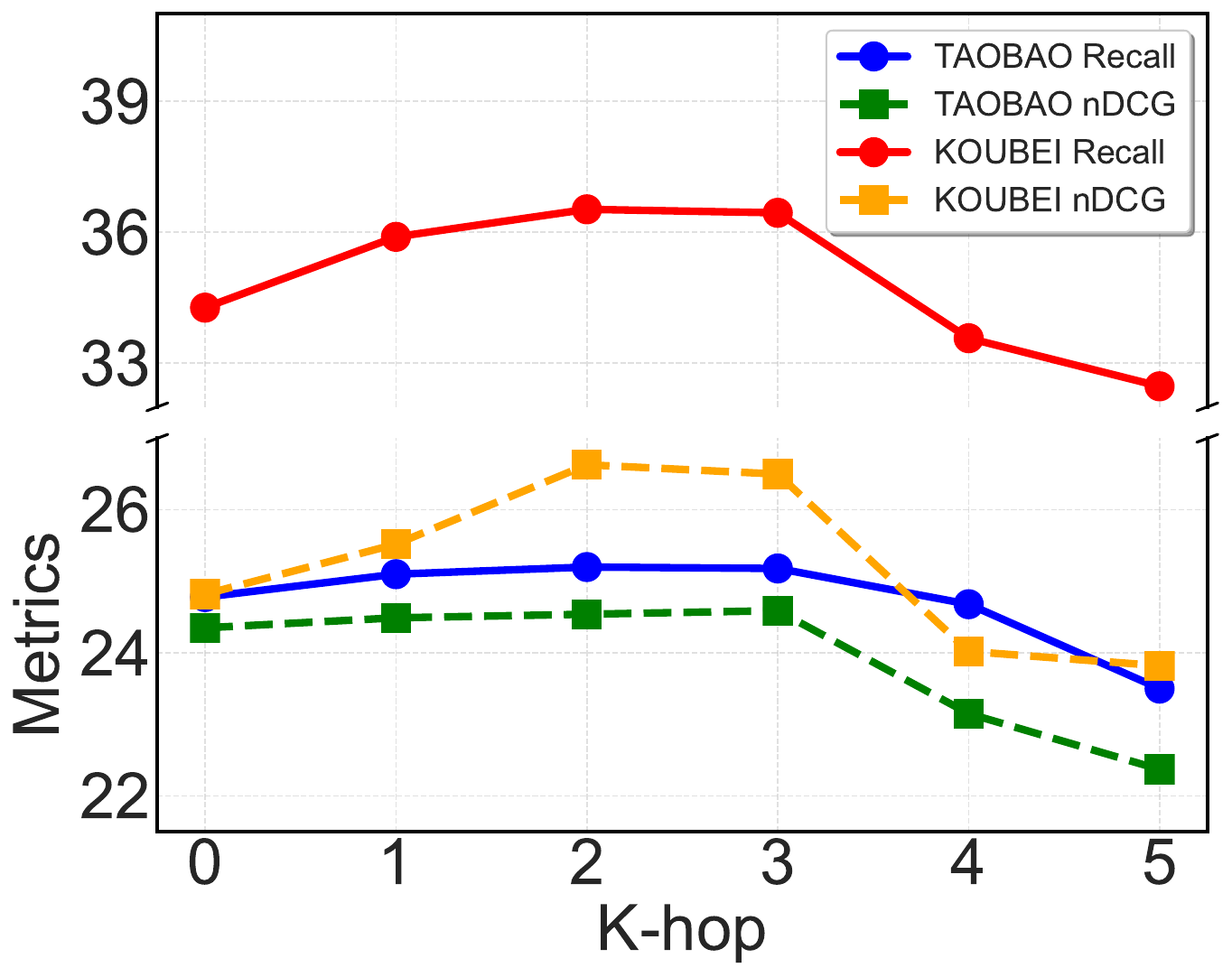}
    \end{minipage}
    }%
    \centering
\caption{Hyper-parameter study results of $M$ and $k$.}
\label{fig:topM_khop}
\vspace{-0.3cm}
\end{figure}

\subsection{Hyperparameter Sensitivity}

We investigate the effect of two key hyperparameters: the number of retrieved subgraphs Top-$M$ and neighborhood depth $k$-hop, as shown in Figure~\ref{fig:topM_khop}. Increasing $M$ introduces more retrieved knowledge, which initially enhances performance by providing richer contextual signals. However, excessive retrieval introduces noise and hinders generalization. Similarly, larger $k$ values enable the model to aggregate broader structural context. Yet, excessive neighborhood expansion leads to oversized subgraphs with redundant information, which may overwhelm the graph encoder, reduce representation quality and impair generalization.

\section{Conclusion}

We present TarDGR, a task-aware retrieval-augmented framework for dynamic graph recommendation. 
By integrating a task-aware evaluation mechanism and a graph transformer-based task-aware model, TarDGR adaptively selects and fuses semantically relevant historical subgraphs to enhance representation learning under temporal dynamics. 
This design effectively mitigates generalization degradation caused by temporal shifts between pretraining and fine-tuning stages. Experimental results validate the effectiveness of TarDGR in dynamic graph recommendation tasks.
In the future, our framework can be extended to other specific tasks by adapting the task-aware objective.

\section*{Acknowledgment}
This work is supported by the Natural Science Foundation of Guangdong Province, China (2024A1515110162).
\bibliography{aaai2026}
\appendix
\clearpage

\section{Theoretical Analysis}
\label{appendix:theory_analysis}
In this section, we provide a theoretical analysis to explain the generalization benefits of our task-aware subgraph augmentation mechanism. Specifically, we rigorously demonstrate how task-aware retrieval augmentation TarDGR improves model performance, especially in comparison to non-task-aware baselines \textsc{RAGraph~\cite{jiang2024ragraph}}. Our analysis is grounded in information theory and investigates how incorporating task-relevant subgraphs enhances mutual information between input features and output labels, thereby boosting generalization.

Let us define the following:

\begin{itemize}
    \item $X$: input features (e.g., embeddings of the query graph),
    \item $Y$: output labels (e.g., target item in graph recommendation or predicted score),
    \item $RAG$: generic retrieval augmentation, which retrieves externally related subgraphs based on input $X$,
    \item $RAG_{\text{aware}}$: task-aware retrieval augmentation, which filters $RAG$ to retain only the most task-relevant subgraphs.
\end{itemize}

We analyze the effectiveness of task-aware retrieval through mutual information. Mutual information $I(X; Y)$ measures the dependency between input $X$ and output $Y$; a higher value indicates stronger relevance and potentially better model performance~\cite{mao2021neuron}.

\subsection{Raw Representations vs. Generic RAG}

The mutual information between $X$ and $Y$ is defined as:
\begin{equation}
I(X; Y) = \sum_{x, y} p(x, y) \log \frac{p(x, y)}{p(x) p(y)}.
\end{equation}

In the presence of retrieval augmentation, we consider the joint input $(X, RAG)$, where $RAG$ provides additional context. The mutual information between the augmented input and the label is:
\begin{equation}
I(X, RAG; Y) = \sum_{x, r, y} p(x, r, y) \log \frac{p(x, r, y)}{p(x, r) p(y)}.
\end{equation}

We are interested in the gain in mutual information due to retrieval:
\begin{equation}
\begin{aligned}
&I(X, RAG; Y) - I(X; Y) \\&= \sum_{x, r, y} p(x, r, y) \log \frac{p(x, r, y)}{p(x, r) p(y)} - \sum_{x, y} p(x, y) \log \frac{p(x, y)}{p(x) p(y)} \\
&= \sum_{x, r, y} p(x, r, y) \log \left( \frac{p(x, r, y)}{p(x, r)} \cdot \frac{p(x)}{p(x, y)} \right) \\
&= \sum_{x, r, y} p(x, r, y) \log \frac{p(x, r, y)}{p(r \mid x) p(x, y)} \\
&= \sum_{x, r, y} p(r, y \mid x) p(x) \log \frac{p(r, y \mid x)}{p(r \mid x) p(y \mid x)} \\
&= I(RAG; Y \mid X).
\end{aligned}
\end{equation}

Thus, we have:
\begin{equation}
{I(X, RAG; Y) - I(X; Y) = I(RAG; Y \mid X)}.
\end{equation}

Since $RAG$ is designed to retrieve relevant information for $X$, the conditional mutual information $I(RAG; Y \mid X)$ is generally non-negative. Therefore, the retrieved knowledge increases the information available to model, yielding:
\begin{equation}
{I(X, RAG; Y) \geq I(X; Y)}.
\end{equation}

This confirms that even generic retrieval augmentation (i.e., \textsc{RAGraph}) can enhance model prediction by increasing the dependency between inputs and outputs.

\subsection{Task-Aware vs. Generic RAG}

In this subsection, we compare the mutual information between task-aware retrieval augmentation (\textsc{TarDGR}) and generic retrieval augmentation (\textsc{RAGraph}). Specifically, we analyze how selectively filtering retrieved subgraphs based on task relevance enhances the mutual information between the input and the output, while suppressing irrelevant noise.

Let $RAG_{\text{aware}}$ denote the task-aware version of retrieval augmentation, which retains only the most task-relevant subgraphs from the generic retrieval $RAG$. Our goal is to compare the following two mutual information quantities:
\begin{align}
\small
I(X, RAG_{\text{aware}}; Y) &= \sum_{x, r_{\text{aware}}, y} p(x, r_{\text{aware}}, y) \log \frac{p(x, r_{\text{aware}}, y)}{p(x, r_{\text{aware}}) p(y)}. \\
I(X, RAG; Y) &= \sum_{x, r, y} p(x, r, y) \log \frac{p(x, r, y)}{p(x, r) p(y)}.
\end{align}

We are interested in the difference:
\begin{equation}
I(X, RAG_{\text{aware}}; Y) - I(X, RAG; Y).
\end{equation}
By expanding the definitions and aligning the summation space:
\begin{equation}
\begin{aligned}
&I(X, RAG_{\text{aware}}; Y) - I(X, RAG; Y) \\
&= \sum_{x, r_{\text{aware}}, y} p(x, r_{\text{aware}}, y) \log \frac{p(x, r_{\text{aware}}, y)}{p(x, r_{\text{aware}}) p(y)}
\\&- \sum_{x, r_{\text{aware}}, y} p(x, r_{\text{aware}}, y) \log \frac{p(x, r, y)}{p(x, r) p(y)} \\
&= \sum_{x, r_{\text{aware}}, y} p(x, r_{\text{aware}}, y) \log \left(
\frac{p(x, r_{\text{aware}}, y)}{p(x, r_{\text{aware}})} \cdot \frac{p(x)}{p(x, r, y)}
\right) \\
&= \sum_{x, r_{\text{aware}}, y} p(x, r_{\text{aware}}, y) \log \frac{p(x, r_{\text{aware}}, y)}{p(r_{\text{aware}} \mid x) p(x, y)}.
\end{aligned}
\end{equation}

This simplifies to:
\begin{equation}
\begin{aligned}
&I(X, RAG_{\text{aware}}; Y) - I(X, RAG; Y) \\
&= I(RAG_{\text{aware}}; Y \mid X) - I(RAG; Y \mid X).
\end{aligned}
\end{equation}

Since task-aware retrieval explicitly filters out irrelevant or noisy subgraphs, the conditional mutual information is strictly improved:
\begin{equation}
{I(RAG_{\text{aware}}; Y \mid X) \geq I(RAG; Y \mid X)}.
\end{equation}

This inequality confirms that: {Task-aware retrieval augmentation is superior to generic retrieval augmentation}. By removing task-irrelevant noise, the model focuses on more discriminative and relevant signals for the downstream objective.

\subsection*{Summary of Theoretical Findings}

Our analysis leads to the following conclusions:
\begin{enumerate}[leftmargin=*]
    \item \textbf{Generic retrieval augmentation improves over raw inputs}: Introducing $RAG$ increases mutual information with the label, i.e., $I(X, RAG; Y) > I(X; Y)$, resulting in improved prediction performance.
    \item \textbf{Task-aware retrieval augmentation improves over generic retrieval}: Selective enhancement via $RAG_{\text{aware}}$ further increases task-relevant information and reduces irrelevant noise, leading to $I(X, RAG_{\text{aware}}; Y) > I(X, RAG; Y)$ and better task alignment.
\end{enumerate}

\section{Further Experiment Details}
\label{appendix:exp_details}
\subsection{Datasets Statics}
\label{appendix: datasets}

\begin{table}[]
    \centering
	\resizebox{\linewidth}{!}{
    \begin{tabular}{lcccc}
    \toprule
    Statistics & TAOBAO & KOUBEI & AMAZON & \\
    \midrule
    \# All Nodes & 204,168 & 221,366 & 238,735 & \\
    \# Users & 117,450 & 119,962 & 131,707 & \\
    \# Items & 86,718 & 101,404 & 107,028 & \\
    \# Interactions & 8,795,404 & 3,986,609 & 876,237 & \\
    \# Density & 8.6e-4 & 3.3e-4 & 6.2e-5 &\\
    \# Snapshot Granularity & daily & weekly & weekly & \\
    \# Dataset Partition & Snapshot & Snapshot & Snapshot & \\ 
    \bottomrule
    \end{tabular}
    }
    \caption{Statistics of the experimental datasets.}
    \label{table:dataset statistics}
\end{table}

We utilize three publicly available datasets that represent diverse real-world dynamic recommendation scenarios:
\begin{itemize}[leftmargin=*]
\item\textbf{TAOBAO } contains implicit feedback collected over 10 consecutive days from Taobao.com, a major Chinese e-commerce platform. It is used for edge classification and includes 204,168 nodes and 8,795,404 interactions, with a graph density of $8.6 \times 10^{-4}$. The data is segmented into daily snapshots over a span of 10 days, with the first 5 days used for pre-training and the remaining 5 days for fine-tuning and prediction.
\item\textbf{KOUBEI } sourced from Koubei, a location-based service within Alipay, this dataset records 9 weeks of user interactions with nearby stores. It comprises 221,366 nodes and 3,986,609 interactions, with a density of $3.3 \times 10^{-4}$, and is also used for edge classification. The dataset covers 9 weeks, with 4 weeks for pre-training and 5 weeks for fine-tuning and evaluation.
\item\textbf{AMAZON } includes 13 weeks of product review interactions from Amazon. It is used for edge classification, containing 238,735 nodes and 876,237 interactions, with a graph density of $6.2 \times 10^{-5}$. The data spans 13 weeks, with a 4-week pre-training period followed by 9 weeks for tuning and evaluation.
\end{itemize}
Detailed statistics of the datasets are presented in Table~\ref{table:dataset statistics}. ``All Nodes" indicates the total number of nodes in the dataset. ``Users" and ``Items" represent the number of user and item nodes, respectively, while ``Interactions" refers to the total number of user and item interactions. ``Snapshot Granularity" specifies the temporal resolution used to segment each dataset. In our experiments, dynamic graphs are partitioned into snapshots based on temporal granularity.

\subsection{Baseline Details}
\label{appendix:baselines}
In this subsection, we present the details of baselines.

\begin{itemize}[leftmargin=*]
    \item \textbf{LightGCN}~\cite{he2020lightgcn}: LightGCN simplifies graph convolutional networks for collaborative filtering by focusing only on neighborhood aggregation. It eliminates unnecessary components like feature transformation and nonlinear activation, which are less effective in recommendation tasks. (Code: \url{https://github.com/kuandeng/LightGCN})

    \item \textbf{SGL}~\cite{sgl}: 
    SGL enhances recommendation tasks by adding an auxiliary self-supervised task that reinforces node representation learning through self-discrimination. (Code: \url{https://github.com/wujcan/SGL-Torch})
    
    \item \textbf{MixGCF}~\cite{huang2021mixgcf}: MixGCF is a negative sampling method for GNN-based recommender systems that synthesizes hard negatives by aggregating embeddings from different layers of raw negatives' neighborhoods using a hop mixing technique. (Code: \url{https://github.com/Wu-Xi/SimGCL-MixGCF})
    
    \item \textbf{SimGCL}~\cite{simgcl}: SimGCL is a simple contrastive learning method for recommendation systems that replaces graph augmentations with uniform noise in the embedding space to create contrastive views. It improves recommendation accuracy and training efficiency by regulating the uniformity of learned representations.
 (Code: \url{https://github.com/Wu-Xi/SimGCL-MixGCF})

    \item \textbf{GraphPrompt}~\cite{graphprompt}: GraphPrompt is a novel pre-training and prompting framework for graphs, designed to bridge the gap between pretraining and downstream tasks. It unifies different tasks into a common template and introduces a learnable task-specific prompt vector to guide each task in utilizing the pretrained model effectively. (Code: \url{https://github.com/Starlien95/GraphPrompt})
    
    \item \textbf{EvolveGCN}~\cite{pareja2020evolvegcn}: EvolveGCN addresses the dynamism of graph sequences by utilizing an RNN to adapt the parameters of the Graph Convolutional Network (GCN) over time. The GCN parameters can be either hidden states (referred to as the -H variant) or inputs of a recurrent architecture (referred to as the -O variant). (Code: \url{https://github.com/IBM/EvolveGCN})

    \item \textbf{ROLAND}~\cite{you2022roland}: ROLAND utilizes a meta-learning approach to update previously learned embeddings for re-initialization. These updated embeddings are then fused with layer-wise hidden states of GNN. (Code: \url{https://github.com/snap-stanford/roland})
    
    \item \textbf{GPF}~\cite{fang2023universal}: GPF introduces prompts within the feature space of the graph, thereby establishing a general approach for tuning prompts in any pre-trained graph neural networks. (Code: \url{https://github.com/LuckyTiger123/GPF})
    \item \textbf{GraphPro}~\cite{yang2024graphpro}: GraphPro enhances GNN-based recommenders by integrating dynamic graph pre-training with temporal and graph-structural prompts, enabling adaptive, accurate recommendations in evolving environments without continuous retraining. (Code: \url{https://github.com/HKUDS/GraphPro})
    
    \item \textbf{PRODIGY}~\cite{mishchenko2023prodigy}: PRODIGY is a pretraining framework that enables in-context learning over graphs. By introducing a prompt graph representation and a graph neural network architecture, PRODIGY allows pretrained models to perform downstream classification tasks on unseen graphs without additional fine-tuning. (Code: \url{https://github.com/snap-stanford/prodigy})
    
    \item \textbf{RAGRAPH}~\cite{jiang2024ragraph}: RAGRAPH introduces a general retrieval-augmented framework for graph neural networks, enabling plug-and-play integration of retrieved subgraph contexts into GNNs without requiring architectural changes. (Code: \url{https://github.com/Artessay/RAGraph})
    
\end{itemize}

\subsection{Evaluation Metrics}
\label{appendix:evaluation-metrics}

Following established practice~\cite{sgl, he2020lightgcn}, we assess the recommendation performance using two widely adopted ranking metrics: Recall@k and NDCG@k, which measure retrieval accuracy and ranking quality, respectively. Let $rel_{ij} \in \{0, 1\}$ denote the binary relevance indicator for the $j$-th predicted item in the top-$k$ list for node $v_i$, where $rel_{ij} = 1$ if the predicted item is a true positive link, and $0$ otherwise. The set of top-$k$ predictions for node $v_i$ is denoted by $\{rel_{ij}\}_{j=1}^k$.

{Recall@k} quantifies the proportion of relevant (ground-truth) items successfully retrieved within the top-$k$ predictions. It is defined as:
\begin{equation}
\text{Recall@}k = \frac{1}{n} \sum_{i=1}^n 
\sum_{j=1}^k \frac{rel_{ij}}{\sum \mathbb{I}(A[i,:] > 0)},
\label{eq:recall}
\end{equation}
where $\mathbb{I}(\cdot)$ is the indicator function, and $A[i,:]$ denotes the ground-truth interaction row for node $v_i$.

{NDCG@k} (Normalized Discounted Cumulative Gain) accounts for both the relevance and the ranking position of the retrieved items. It is computed by first calculating DCG@k as:
\begin{equation}
\text{DCG@}k = \frac{1}{n} \sum_{i=1}^n 
\sum_{j=1}^k \frac{rel_{ij}}{\log_2 (j + 1)},
\label{eq:dcg}
\end{equation}

\subsection{Implementation Details}
\label{appendix:setting-parameter}
The experiments in this paper are conducted using Python 3.9 with PyTorch version 1.13.1. The system is running on Ubuntu 22.04, equipped with an Intel(R) Xeon(R) Platinum CPU.
All experiments are conducted on four NVIDIA GeForce RTX 4090 GPUs, each with 32GB of memory. For LightGCN, SGL, MixGCF, SimGCL and GraphPro, we employ a 3-layer GNN architecture and set the hidden dimension as 64. The hyperparameter $\mu$ is fixed at $1e-4$, $\rho$ is searched over the set $[0.3,\, 0.6,\, 0.9]$, and $\lambda$ is selected from $[0.1,\, 0.2,\, 0.3,\, 0.4,\, 0.5]$ via grid search. For GraphPro, PRODIGY and RAGRAPH, we follow the settings from RAGRAPH~\cite{jiang2024ragraph}. 
In our retrieval framework, $k$-hop is set to 2, $topK$ is set to 5, $topM$ is selected from $[1,\, 2,\, 3]$, edge dropout rate is set to 0.5, learning rate is set to $1e-3$.

All models are initially pre-trained on historical graph snapshots and subsequently fine-tuned and evaluated on future snapshots, ensuring temporal consistency and realistic generalization assessment. For fair comparisons, for methods employing PRODIGY,RAGRAPH and TarDGR, we fine-tune models using the training set while retrieving the resource graph to prevent information leakage and over-fitting; when testing, we retrieve the combined training and resource graphs. For other methods, fine-tuning was directly performed on the combined train and resource set for fairness. We utilize the pre-trained process dynamic graph dataset as the resource pool. For retrieval-based methods, we evaluate two settings: (1) {Non-Fine-tuned (NF)}, which applies plug-and-play retrieval augmentation without further fine-tuning on the training set, and (2) {Fine-tuned (FT)}, which employs prompt tuning on the training set to adapt the retrieved content.

\section{Algorithms}
In this section, we provide a detailed description of the algorithms of Task-Aware Evaluation Mechanism and Graph Transformer-based Task-Aware
Recommendation.

\begin{algorithm}[h]
    \caption{Task-Aware Evaluation Mechanism}
    \label{alg:algorithm_TAE}
    \LinesNumbered
    \KwIn{Pre-trained GNN recommendation model $f^{\theta}_{\text{pre}}$, query subgraph $G(v_q)$, sampled candidate historical subgraph $\{G(v_r)\}_{r=1}^{R_\text{sample}}$, positive subgraphs $\{ G(v_q)_i^+ \}_{i=1}^N$.}
    \KwOut{Task relevance score $C_r$,task-aware dataset $\mathcal{D}_{\text{aware}}$.}
    
    \ForEach{positive subgraph $G(v_q)^+_i \in \{ G(v_q)_i^+ \}_{i=1}^N$}{
        $h_q \gets f_{\text{pre}}\big(G(v_q)\big)$ \tcp*{encode query}\
        $h_{q_i} \gets f_{\text{pre}}\big(G(v_q)^+_i\big)$; 
    }
    $\overline{\textsc{Sim}}_{\text{before}} \gets \frac{1}{N^+} \sum_{i=1}^{N^+} \textsc{Cos}\left(h_q, h_{q_i}\right)$;

    \ForEach{candidate subgraph $G(v_r) \in \{ G(v_r) \}_{r=1}^{R_{\text{sample}}}$}{
        $\tilde{G(v_q)} \gets G(v_q) \oplus G(v_r)$  \tcp*{link}   \
        $h_{\tilde{q}} \gets f_{\text{fuse}}\big(f_{\text{pre}}\big(\tilde{G(v_q)}\big)\big)$ \
        $\overline{\textsc{Sim}}_{\text{after}} \gets \frac{1}{N^+} \sum_{i=1}^{N^+} \textsc{Cos}\left(h_{\tilde{q}}, h_{q_i}\right)$;\
    $\Delta \textsc{Rel} \leftarrow \overline{\textsc{Sim}}_{\text{after}} - \overline{\textsc{Sim}}_{\text{before}}$ \tcp*{score}\
    \eIf{$\Delta \textsc{Rel} > 0$}{
        $C_r \leftarrow$ beneficial sample;
    }{
        \eIf{$\Delta \textsc{Rel} < 0$}{
            $C_r \leftarrow$ harmful sample;
        }{
            $C_r \leftarrow$ irrelevant sample;
        }
    }
    Store triplet $(G(v_q), G(v_r), C_r)$ in dataset $\mathcal{D}_{\text{aware}}$;\
    }
    \Return Task relevance scores $C_r$, dataset $\mathcal{D}_{\text{aware}}$
\end{algorithm}

\subsection{Task-Aware Evaluation Mechanism}

In Algorithm~\ref{alg:algorithm_TAE} , we present a task-aware evaluation mechanism for candidate subgraphs. Initially, we define the necessary inputs, which include the pre-trained GNN recommendation model $f^{\theta}_{\text{pre}}$, the query subgraph $G(v_q)$, sampled candidate historical subgraphs $\{G(v_r)\}_{r=1}^{R_\text{sample}}$, and the set of positive historical subgraphs $\{ G(v_q)^+ \}_{r=1}^{N}$. The outputs are the task relevance scores $C_r$ for the candidate subgraphs and the task-aware dataset $\mathcal{D}_{\text{aware}}$.

In lines 1-3, the process begins with the encoding of the query subgraph and each positive subgraph. Using the pre-trained model $f_{\text{pre}}$, we generate embeddings for the query subgraph $h_q$ and each positive subgraph $h_{q_i}$. The cosine similarity $\overline{\textsc{Sim}}_{\text{before}}$ between the query subgraph and the positive subgraphs is computed in line 4, serving as a baseline for relevance.

Next, in lines 5-7, for each candidate subgraph $G(v_r)$, we link it with the query subgraph to form the combined graph $\tilde{G(v_q)}$. We then apply graph convolution to obtain the fused representation $h_{\tilde{q}}$. The updated cosine similarity $\overline{\textsc{Sim}}_{\text{after}}$ between the fused query-candidate representation and the positive subgraphs is calculated in line 8.

The change in similarity $\Delta \textsc{Rel}$, representing the effect of the candidate subgraph on relevance, is computed in line 9. Based on this shift, the candidate subgraph is classified as beneficial, harmful, or irrelevant in lines 10-16.

Finally, in line 17, the triplet consisting of the query subgraph, the candidate subgraph, and the task relevance score $C_r$ is stored in the task-aware dataset $\mathcal{D}_{\text{aware}}$. The algorithm concludes by returning the task relevance score $C_r$ and the dataset $\mathcal{D}_{\text{aware}}$ in line 18.

\begin{algorithm}[h]
\caption{Graph Transformer-based Task-Aware Recommendation}
\label{alg:graph_transformer_retrieval}
\LinesNumbered
\KwIn{
Query subgraph $G(v_q)$, 
Resource subgraphs $\mathcal{G}_r$, 
Dynamic encoder $\textsc{forward}(\cdot)$, Retrieval size $K$, Augmentation size $M$.}

\KwOut{Query recommendation representation $\tilde{h}_q$.}

$h_t \leftarrow \textsc{forward}(h_{t-1}; G_{t-1})$;\;
\SetKwProg{Fn}{Function}{:}{end}

\Fn{\textsc{DistanceRetrieval}($\mathcal{G}_r$, $G(v_q)$)}{
    \ForEach{resource subgraph $G(v_r) \in \mathcal{G}_r$}{
        $h_r \leftarrow  \text{Eq.(\ref{subgraph_rep})} $ \tcp*{GConv embed.}
    }\
    $h_q \leftarrow \sum_{l=0}^{L} \textsc{GConv}(h_t^q, G(v_q))$\;
    $\mathcal{G}^K(v_q) \leftarrow \textsc{TopK}_{\text{search}}(G(v_q), G(v_r))$\;
    \Return{candidate subgraphs $\mathcal{G}^K(v_q)$}
}

\Fn{\textsc{TaskAwareEnc}($\mathcal{G}^K(v_q)$, $G(v_q)$ ,$\mathcal{A}_s$)}{
    \ForEach{candidate subgraph $G(v_i) \in \mathcal{G}^K(v_q)$}{
        Encode $G(v_q)$ and $G(v_i)$ to obtain $h_q, h_i$\;
        $h \leftarrow [h_q \,\|\, h_i]$\; 
        $h_{\text{pos}} \leftarrow h + P$\;
        $\text{Attn} \leftarrow \textsc{MultiHeadAttn}(h_{\text{pos}})$\;
        $h_{\text{sem}} \leftarrow \text{Eq.(\ref{eq:sem_concat})}$ \tcp*{semantic emb.}\
        $h_{\text{hid}} \leftarrow h_{\text{pos}}W + b$\;
        $h_{pos}^\text{Attn} \leftarrow \textsc{MultiLayerAttn}(h_{\text{pos}})$\;
        $h_{\text{ffn}}\leftarrow \textsc{FFN}(h_{pos}^\text{Attn})$\;
        $h_{\text{str}} \leftarrow \text{Eq.(\ref{eq:str_adj_pro})}$ \tcp*{structure emb.}\
        $h_{\text{task}} \leftarrow [h_{\text{sem}} \| h_{\text{str}}]$\;
        $s_i \leftarrow\mathcal{S}_\psi(h_{\text{task}})$ \tcp*{relevance score}
    }\
    \Return{task-aware relevance score $\{{s^{(i)}}_{i=1}^K\}$ }
}

Select top-$M$ candidates $\mathcal{G}_m =\{G(v_m)\}_{m=1}^M$ with highest scores $\{{s^{(i)}}_{i=1}^K\}$\;
\ForEach{task-relevant subgraph $G(v_m) \in \mathcal{G}_m$}{
    $h_m \leftarrow \text{Eq.(\ref{eq:intra_agg})}$ \tcp*{intra aggregate}
}\
$H_{\text{rag}} \leftarrow \text{Eq.(\ref{eq:soft_agg})}$ \tcp*{inter aggregate}\
$\tilde{h}_q \leftarrow \text{Eq.(\ref{eq:fuse_agg})}$ \tcp*{task-enhanced emb.}\
\Return{$\tilde{h}_q$}

\end{algorithm}

\subsection{Graph Transformer-based Task-Aware Recommendation}

Algorithm~\ref{alg:graph_transformer_retrieval} outlines the procedure of the proposed Graph Transformer-based Task-Aware Recommendation framework. Given a query subgraph $G(v_q)$, a resource subgraph pool $\mathcal{G}_r$, a pretrained dynamic encoder $\textsc{forward}(\cdot)$, and hyperparameters $K$ and $M$ for retrieval and augmentation, the algorithm outputs the enhanced query representation $\tilde{h}_q$ for downstream recommendation.

Line 1 initializes temporally contextualized node embeddings $h_t$ via the dynamic encoder applied to the graph snapshot at time $t{-}1$, capturing historical user-item dynamics.

In lines 2–7, each resource subgraph $G(v_r) \in \mathcal{G}_r$ is encoded via $L$-layer graph convolution to obtain subgraph-level representations $h_r$. The query subgraph is similarly encoded as $h_q$. A distance-based retrieval mechanism selects the top-$K$ most similar candidate subgraphs, forming the coarse retrieval set $\mathcal{G}^K(v_q)$.

Lines 8–21 performs pairwise task-aware encoding between $G(v_q)$ and each $G(v_i) \in \mathcal{G}^K(v_q)$. The concatenated embeddings $[h_q ,|, h_i]$ are augmented with positional encodings and passed through a multi-head self-attention module, yielding semantic representations $h_{\text{sem}}$. Parallelly, the same input undergoes latent projection, structural attention encoding with residual normalization, and feedforward transformation, followed by adjacency-aware propagation to derive structural representations $h_{\text{str}}$. The combined representation $h_{\text{task}} = [h_{\text{sem}} ,|, h_{\text{str}}]$ is scored by a learnable relevance function $\mathcal{S}_\psi(\cdot)$ to estimate task-specific importance $s_i$.

In lines 22–26, the top-$M$ most relevant subgraphs are selected. Each is re-encoded via intra-subgraph aggregation to obtain $h_m^i$, and a weighted combination using confidence scores $\alpha_i$ produces the retrieved knowledge representation $H_{\text{rag}}$. The final task-enhanced query embedding is computed via gated residual fusion: $\tilde{h}_q = \beta h_q + (1 - \beta) H_{\text{rag}}$, as shown in line 26. The algorithm concludes by returning the final task-aware enhanced query recommendation representation $\tilde{h}_q$ in line 27.

\section{Data Ethics Statement}
To evaluate the efficacy of this
work, we conducted experiments that only use publicly available
datasets, namely, TAOBAO, KOUBEI and AMAZON in accordance to their usage terms
and conditions if any. We further declare that no personally identifiable information was
used, and no human or animal subject was involved in this research.

\section{Complexity Analysis}

To assess the scalability of TarDGR, we provide a complexity analysis. TarDGR adopts a strategy of leveraging offline computational investment to enhance online recommendation accuracy. Given the small retrieval size $K$ and fusion size $M$, the framework ensures controllable complexity for large-scale applications.

\noindent\textbf{Offline Stage.} The offline phase primarily consists of the task-aware evaluation and model pre-training.
For the \textit{Task-Aware Evaluation}, given $N_Q$ query subgraphs and $R_s$ candidates per query, the process involves graph fusion and an $L$-layer GNN propagation. With $N_s$ nodes, $E_s$ edges, and embedding dimension $d$, the time complexity is $\mathcal{O}(N_Q \cdot R_s \cdot L \cdot E_s \cdot d)$.
For \textit{Pre-training}, the computation is dominated by the multi-head self-attention mechanism in the Graph Transformer. With a dataset size of $|D_{aware}|$, the complexity is $\mathcal{O}(|D_{aware}| \cdot N_s^2 \cdot d)$.
In terms of \textit{Space Complexity}, the memory consumption is dominated by storing the task-aware dataset and model parameters ($\theta_{gnn}, \theta_r$), resulting in $\mathcal{O}(\theta_{gnn} + \theta_r + N_Q \cdot R_s \cdot (N_s + E_s + N_s d))$.

\noindent\textbf{Online Stage.} The online inference involves retrieval, re-ranking, and fusion. Let $|G_R|$ denote the resource library size.
The \textit{Time Complexity} consists of three parts: (1) FAISS vector indexing with $\mathcal{O}(d \cdot \log |G_R|)$; (2) Task-aware re-ranking for top-$K$ candidates with $\mathcal{O}(K \cdot N_s^2 \cdot d)$; and (3) Subgraph fusion involving $M$ GNN forward passes with $\mathcal{O}(M \cdot L \cdot E_s \cdot d)$.
The \textit{Space Complexity} during inference requires storing the resource library and temporary attention matrices, formulated as $\mathcal{O}(\theta_{gnn} + \theta_r + |G_R| (N_s + E_s + N_s d) + K N_s^2)$.

\section{Ablation Study on BiSCL}
To investigate the effectiveness of the Bi-Level Supervised Correlation Loss (BiSCL), we conducted an ablation study comparing TarDGR with three variants:
 \textbf{w/o BiSCL}: The task-aware model is randomly initialized without pre-training;
 \textbf{w/o ocl}: Pre-trained using only the Magnitude Fitting Loss ($\mathcal{L}_{mtl}$); and
 \textbf{w/o mtl}: Pre-trained using only the Ordinal Contrast Loss ($\mathcal{L}_{ocl}$).
\begin{table}[h]
    \centering
    \caption{Ablation study of BiSCL on Koubei and Amazon.}
    \label{tab:biscl_ablation}
    \setlength{\tabcolsep}{5pt}
    \begin{tabular}{lcccc}
        \toprule
        \multirow{2}{*}{\textbf{Method}} & \multicolumn{2}{c}{\textbf{Koubei}} & \multicolumn{2}{c}{\textbf{Amazon}} \\
        \cmidrule(lr){2-3} \cmidrule(lr){4-5}
         & Recall & nDCG & Recall & nDCG \\
        \midrule
        w/o BiSCL & 35.62 & 25.98 & 19.28 & 09.53 \\
        w/o ocl & 36.18 & 26.36 & 19.40 & 09.59 \\
        w/o mtl & 36.27 & 26.45 & 19.46 & 09.64 \\
        \textbf{TarDGR} & \textbf{36.52} & \textbf{26.63} & \textbf{19.56} & \textbf{09.70} \\
        \bottomrule
    \end{tabular}
\end{table}

The results in Table \ref{tab:biscl_ablation} demonstrate that removing BiSCL entirely leads to the most significant performance degradation, highlighting the necessity of task-aware supervision. Furthermore, both the magnitude fitting ($\mathcal{L}_{mtl}$) and ordinal contrast ($\mathcal{L}_{ocl}$) objectives contribute independently to the performance. The full TarDGR model achieves the best results, indicating that jointly optimizing for absolute relevance scores and relative ranking order enables the model to effectively distinguish task-beneficial subgraphs.



\end{document}